\documentstyle[preprint,prd,aps,graphicx]{revtex}
\input BoxedEPS
\SetRokickiEPSFSpecial  
\HideDisplacementBoxes 

\def\rd#1{\mathop{{\rm d}#1}}

\newcommand{\ignore}[1]{} 

\tighten

\begin{document}
\draft
\preprint{MIT-CTP-3219, JLAB-THY-02-06}
\title{Moments of Nucleon Light Cone Quark Distributions \\
       Calculated in Full  Lattice QCD\\}
\author{LHPC and SESAM Collaborations}
\author{D. Dolgov, R. Brower, S. Capitani\footnote{present address
DESY/Zeuthen}, P. Dreher, J. W. Negele, A. Pochinsky, D.  B. Renner}
\address{Center for Theoretical Physics, Laboratory for Nuclear Science, \\
Massachusetts Institute of Technology, \\
77 Massachusetts Avenue, Cambridge, Massachusetts 02139\\}
\author{N. Eicker, Th. Lippert, K. Schilling}
\address{Department of Physics, University of Wuppertal, D-42097 Wuppertal, 
Germany\\}
\author{R. G. Edwards}
\address{Jefferson Lab, 12000 Jefferson Avenue, MS 12H2,
Newport News, Virginia 23606\\}
\author{U. M. Heller}
\address{CSIT, Florida State University, Tallahassee, Florida 32306\\}
\date{January 28, 2002}
\maketitle
\begin{abstract}
Moments of the quark density, helicity, and transversity distributions are
calculated in unquenched lattice QCD. Calculations of proton matrix elements of
operators corresponding to these moments through the operator product 
expansion have been performed on
$16^3
\times 32$ lattices for Wilson  fermions at $\beta = 5.6$ using
configurations from the SESAM collaboration and at $\beta = 5.5$ using
configurations from SCRI. One-loop perturbative renormalization
corrections are included. 
At quark masses  accessible in present calculations, there is no statistically
significant difference between quenched and full QCD results, indicating that the
contributions of quark-antiquark excitations from the Dirac Sea are small.  Close
agreement between calculations with cooled configurations containing
essentially only instantons and the full gluon configurations indicates that
quark zero modes associated with instantons play a dominant role.  Naive
linear extrapolation of the full QCD calculation to the physical pion mass
yields results inconsistent with experiment.  Extrapolation to the chiral
limit including  the physics of the pion cloud can resolve this discrepancy
and the requirements for a definitive chiral extrapolation are described.  
\end{abstract}


%
%

\section{Introduction}

The quest to understand the matter of which our universe is composed will
remain fundamentally incomplete until we understand how the quark and gluon
structure of the nucleon arises from QCD.  The nucleon has many remarkable
properties.   Because of confinement, the
quark-gluon structure of hadrons differs essentially from that of any other
known composite systems. Gluons in QCD are essential dynamical 
degrees of freedom, unlike the
boson fields in atoms and nuclei which may be subsumed into a two-body
interaction thereby reducing these systems to purely fermionic degrees of
freedom.  Almost all of the  mass and
approximately half of the momentum and angular momentum of a nucleon
arises from gluons. Indeed, even the net spin 1/2 arises from a rich and
complicated combination of orbital and intrinsic angular momentum of the
quark and gluon fields. Since the usual analytic tools of theoretical physics have
proven inadequate to solve nonperturbative QCD, the only known way to solve,
rather than model, QCD is numerical solution of lattice field theory.  Hence, the
ultimate goal of this work is to use lattice QCD to understand the structure of
the nucleon. This understanding has two important but distinct aspects.

One aspect of using lattice QCD to understand nucleon structure is the
quantitative {\it ab initio} calculation of experimental observables. Since the
experimental discovery of quarks in the nucleon over a quarter of a century ago,
there has been a huge investment internationally in using high energy
scattering to measure the light cone distribution of quarks and gluons in the
nucleon. As a result of several decades of experimental effort at SLAC, Fermilab,
CERN, and DESY, we now have a detailed knowledge of the quark density
and helicity distributions and  of the gluon
distribution
\cite{bib-cteq,bib-grv,bib-mrs,bib-grsv,bib-gs,bib-smc1,bib-smc2,bib-hermes}.
In addition, major new experiments are being planned at these facilities as well
as at Jefferson Lab and RHIC to map out the quark and gluon structure of the
nucleon in even more detail.  Now that the techniques of lattice field theory and
computer technology have developed to the point that it will be possible to
solve QCD with a precision comparable to experimental measurements, it is
essential to complement this massive experimental investment with a
commensurate theoretical effort in lattice QCD.  Because deep inelastic lepton
scattering measures correlation functions close to the light cone, structure
functions are intrinsically Minkowski and cannot be calculated directly in lattice
QCD.  However, using the operator product expansion, it is possible to calculate
their moments. In this work, we report the  first calculations of these
moments in full QCD\cite{Dolgov:2000ca}. As will be emphasized below, these
present calculations are subject to significant limitations due to current
computer resources.

The second aspect is understanding the basic mechanisms underlying nucleon
structure - that is, how QCD actually works.  Hence, we seek to use the lattice as
a tool for insight as well as for numbers.  There are several ways lattice
calculations can provide insight. One is to calculate the overlap between a trial
wave function and the exact nucleon wave function to explore the role of
various degrees of freedom and variational parameters.  Thus, this work will
present a prototype variational calculation in which the rms radius of a trial
function is varied.  A second is to study the contributions of different classes of
Feynman diagrams that correspond to separate lattice contributions to operators.
An example is the class of connected diagrams considered in this work. Finally,
since the lattice Monte Carlo calculation stochastically samples gluons
distributed according to the QCD action, one can identify and study the dominant
configurations.  Hence, this work will study the role of instantons and their
associated zero modes in calculating moments of structure functions.

There have been several calculations of moments of structure functions in recent
years, including contributions of connected diagrams  to low moments of
the spin  independent and longitudinal spin dependent structure functions in
quenched lattice QCD
\cite{bib-qcdsf-main,bib-qcdsf-1997,Dolgov:1998js,Dolgov:2000ca}, disconnected
contributions for the  axial and  tensor charge \cite{bib-liu}, and the axial charge
in full as well as quenched QCD \cite{Gusken:1999xy}.  Interestingly, in contrast
to spectroscopy, there are significant discrepancies between these
hadron structure calculations and experiment.  Whereas quenched masses for
light hadrons are typically accurate at the 5\% level, the axial charge is typically
10 - 20 \% low  \cite{bib-qcdsf-tensor2} and the first moment of the spin
averaged structure function is of the order of 50\% high 
\cite{bib-qcdsf-main,bib-qcdsf-1997}.
By the variational principle, we know that an error of order $\epsilon$ in the
wave function only produces an error of order
$\epsilon^2$ in the expectation value of the energy, or mass in this case, so it is
consistent that these other observables should be much better diagnostics than
masses of the errors in the lattice calculation of hadron structure. Hence, we
need to identify and correct the source of these substantial discrepancies.

There are three major limitations in structure calculations imposed by
limitations in computational resources.  The first limitation is errors in
approximating the continuum limit. The approach to the continuum limit has
been studied systematically in quenched QCD by the QCDSF
collaboration\cite{bib-qcdsf-main} using Wilson and  clover improved
actions and extrapolating in inverse coupling
$\beta$. A particularly careful study was carried out for the axial charge 
which showed that even when the $\beta$ dependence is included, there
remains a 10\% discrepancy with experiment\cite{bib-qcdsf-tensor2}. 

The second limitation is the
quenched approximation, which ignores the contribution of dynamical
quark-antiquark excitations in the Dirac sea. It is now known that
whereas full QCD gives a good description of masses of hadrons containing
strange quarks, quenching causes significant discrepancies in masses. Physically,
this is consistent with the fact that in quenched QCD the coupling runs too fast,
producing a potential that becomes too weak at the short distances probed by
the relatively heavy strange quark. Although there is no simple argument
suggesting the sign of the resulting discrepancy in other specific observables, it is
important to see to what extent quenching causes the discrepancy in the axial
charge and the first moment of structure functions.  Thus, a primary goal of this
work is to study the role of the quenched approximation. Although the effects of
the lattice spacing and lattice volume are still significant potential sources of
error, it is meaningful to compare quenched and unquenched calculations on
comparable lattices in physical units to explore the magnitude of errors due to
quenching.  Hence, we will compare full QCD calculations using 
configurations produced by the
SESAM collaboration~\cite{bib-sesam-spectro}
at $\beta = 5.6$ with
quenched calculations on comparable lattices.  Whereas it has been conjectured
that the discrepancy between lattice calculations and experimental moments of
structure functions arises primarily from quenching, one major result of
this work is to show that at the quark masses attainable at present, this
explanation is wrong.

The final limitation is errors
in calculating the large volume, small quark mass limit. Since the lattice volume
must be large enough that the pion Compton wavelength fits well inside and
inversion algorithms become less efficient for light quarks, it
is presently impossible to perform full QCD calculations with dynamical
quark masses corresponding to
physical pions or even to pions sufficiently light that chiral perturbation theory
gives reliable extrapolations. 
Thus, a major uncertainty is the extrapolation from
the relatively heavy quark masses used in lattice calculations to the small
quark mass required to produce the physical pion. This is particularly
worrisome for calculating hadron structure because of the major role
played by the pion cloud in the nucleon. Since a small box and heavy quarks
suppress the pion cloud, it is not surprising that the nucleon magnetic
moment, much of which comes from the pion current, is low or that the
axial charge is too small. Indeed, model estimates
\cite{Leinweber:1998ej,Detmold:2001jb} show that both discrepancies
plausibly arise from omission of the full contribution of the pion cloud in
present lattice calculations.  Hence, a second major result of this work is
to show how an extrapolation incorporating the leading effects of chiral
symmetry can simultaneously resolve the discrepancy in the lowest three
moments of the spin averaged structure function.

The outline of this paper is as follows. Section II provides the background and
defines the operators we evaluate. Perturbative renormalization is discussed in
Section III and technical details of the lattice calculation are given in Section IV. 
Section V presents the results, including the overlap of trial wave
functions with lattice hadron ground states, comparison of observables calculated
with quenched and unquenched configurations to study quenching errors,
comparison of observables calculated with cooled and uncooled  unquenched
configurations to study the role of instantons and their associated zero modes,
and comparison of linear and chiral extrapolations with phenomenology. A
summary and conclusions are given in Section VI. 

\section{Background}

\subsection{Moments of nucleon light cone quark distributions}

By the operator product expansion, moments of the  linear combinations of
quark and antiquark distributions in the proton
\begin{eqnarray}
\langle x^n\rangle_q &=&
\int_0^1 \rd x x^n \bigl(q (x) + (-1)^{n+1}\bar q(x)\bigr)  \label{mom1}\\
\langle x^n\rangle_{\Delta q} &=&
\int_0^1 \rd x x^n \bigl(\Delta q (x) + (-1)^{n}\Delta\bar q(x)\bigr)  \nonumber\\
\langle x^n\rangle_{\delta q} &=& 
\int_0^1 \rd x x^n \bigl(\delta q (x) + (-1)^{n+1}\delta\bar q(x)\bigr) \nonumber,
 \end{eqnarray}
where  the quark density, helicity, and transversity
\cite{bib-jaffe-t1} distributions
\begin{eqnarray}
q & =  & q_\uparrow + q_\downarrow \\
\Delta q & = & q_\uparrow - q_\downarrow \nonumber\\
\delta q & = & q_\top  - q_\bot  \nonumber,
\end{eqnarray}
are related to the following matrix elements of twist-2 operators
\begin{eqnarray}
2\, \langle x^{n-1}\rangle_{ q_r } \, P_{\mu_1} \cdots P_{\mu_n} &\equiv&
\frac{1}{2} \sum_S\langle PS | \left(\frac{i}{2}\right)^{n-1} 
\bar{\psi^r}\gamma_{\{\mu_1}{\stackrel{\,\leftrightarrow}{D}}_{\mu_2} \cdots
{\stackrel{\,\leftrightarrow}{D}}_{\mu_n\}}\psi^r | PS\rangle 
\label{ourmoments}\\
\frac{2}{n+1} \, \langle x^n \rangle_{ \Delta q_r}\, S_{\{\sigma} P_{\mu_1}
\cdots P_{\mu_n\}} &\equiv& - \langle PS | \left(\frac{i}{2}\right)^{n} 
\bar{\psi^r} \gamma_5\gamma_{\{\sigma}
{\stackrel{\,\leftrightarrow}{D}}_{\mu_1} \cdots
{\stackrel{\,\leftrightarrow}{D}}_{\mu_n\}}\psi^r | PS\rangle \nonumber \\
\frac{2}{m_N} \, \langle x^n \rangle_{ \delta q_r}\, S_{[\mu} P_{\{\nu]}
P_{\mu_1} \cdots P_{\mu_n\}} &\equiv& \langle PS | \left(\frac{i}{2}\right)^{n} 
\bar{\psi^r} \gamma_5 \sigma_{\mu \{\nu}
{\stackrel{\,\leftrightarrow}{D}}_{\mu_1} \cdots
{\stackrel{\,\leftrightarrow}{D}}_{\mu_n\}}\psi^r | PS\rangle , \nonumber
\end{eqnarray}
Here, $ {\stackrel{\,\leftrightarrow}{D}} \equiv
{\stackrel{\,\rightarrow}{D}} - {\stackrel{\,\leftarrow}{D}}$, $r$ denotes
the quark flavor, 
$x$ denotes the
momentum fraction carried by the quark, $S^2 = m_N^2$,  $\{\,\} $ and $[\,] $
denote symmetrization and  antisymmetrization respectively, and the mixed
symmetry $[\,\{\,]\,\}$ term is first symmetrized and then antisymmetrized so
that it is written explicitly as
\begin{equation}
O^5_{[\sigma\{\mu_1]\cdots\mu_n\}}\equiv\frac{1}{n+1}
\left(O^5_{\sigma\mu_1\mu_2\ldots\mu_n}-O^5_{\mu_1\sigma\mu2\cdots\mu_n}
+O^5_{\sigma\mu_2\mu_1\cdots\mu_n}
-O^5_{\mu_1\mu_2\sigma\cdots\mu_n}+\cdots\right) .
\end{equation}

We note that the odd
moments 
$\langle x^n\rangle_q$ are obtained from  the spin-independent
structure functions
$F_1$ or
$F_2$  measured in deep inelastic electron or muon 
scattering
\begin{eqnarray}
\int^1_0 dx \, x^{n-1} \, F_1(x,Q^2) &=& \frac{1}{2} C_n^v (Q^2/\mu^2) \, 
\sum_r \, e^2_r \, \langle x^{n-1}\rangle_{ q_r } (\mu) \label{dis-momsum2}\\
\int^1_0 dx \, x^{n-2} \, F_2(x,Q^2) &=& C_n^v (Q^2/\mu^2) \,
\sum_r \, e^2_r \, \langle x^{n-1}\rangle_{ q_r } (\mu)  , \nonumber
\end{eqnarray}
 and even moments of $\langle x^n\rangle_{\Delta q} $ are determined
from the spin-dependent structure function $g_1$
\begin{equation}
\int^1_0 dx \, x^n \, g_1(x,Q^2) = \frac{1}{4} C_n^a (Q^2/\mu^2) \, 
\sum_r \, e^2_r \, 2 \langle x^n \rangle_{ \Delta q_r} (\mu) 
\label{dis-momsum1}, 
\end{equation}
where  $e_r$ is the
quark's electric charge, and $C_n$ denotes the Wilson coefficient.
Note that the moments $\langle x^n\rangle_q$ and $\langle x^n\rangle_{\Delta
q} $ are proportional to the quantities $v_{n+1}$ and
$a_n$ defined in Ref.~\cite{bib-qcdsf-main}
\begin{eqnarray}
\langle x^n \rangle_q &=& v^{(q)}_{n+1} \label{res-va2pdf}\\
\langle x^n \rangle_{ \Delta q} &=& \frac{1}{2}\,a^{(q)}_{n}\nonumber .
\end{eqnarray}

In addition, the two spin-dependent structure functions $g_1$ and $g_2$
also determine the quantity
$d_n$
\begin{equation} \label{eqn_dn}
\frac{1}{n+1} \, d^r_n \, S_{[\sigma} P_{\{\mu_1]}\cdots P_{\mu_n\}}
\equiv - \langle PS | \left(\frac{i}{2}\right)^{n} \bar{\psi^r}
\gamma_5\gamma_{[\sigma}{\stackrel{\,\leftrightarrow}{D}}_{\{\mu_1]} \cdots
{\stackrel{\,\leftrightarrow}{D}}_{\mu_n\}}\psi^r | PS\rangle \nonumber 
\end{equation}
which is a twist-three operator and does not have a simple interpretation 
in terms of parton distribution functions \cite{Wandzura:qf}. 
However, since with Wilson fermions,
$\gamma_5\gamma_{[\sigma}{\stackrel{\,\leftrightarrow}{D}}_{\{\mu_1]}
\cdots {\stackrel{\,\leftrightarrow}{D}}_{\mu_n\}}$ mixes with the lower
dimension operator
$\frac{1}{a} \gamma_5\gamma_{[\sigma}\gamma_{\{\mu_1]} \cdots
{\stackrel{\,\leftrightarrow}{D}}_{\mu_n\}}$, it is  not possible to compare
with phenomenological results using the perturbative renormalization
constants and mixing coefficients calculated in this work. 
Either nonperturbative
renormalization is required \cite{bib-qcdsf-nonpert-renorm} as has been
carried
out in \cite{Gockeler:2000ja},  or the operators
need to be recalculated with overlap fermions\cite{Capitani:2000aq} or
some alternative formulation for which mixing with lower dimension
operators does not occur .

Even moments  $\langle
x^n\rangle_q$ are obtained from deep inelastic neutrino scattering, and in addition, a
variety of other processes have contributed to what is now a detailed empirical
knowledge of the quark and antiquark distributions in the nucleon. 
Hence, we will
subsequently compare our  results with moments calculated from the 
CTEQ, GRV, MRS, GRSV, and GS and global fits to the world supply of
data\cite{bib-cteq,bib-grv,bib-mrs,bib-grsv,bib-gs}.
The moments of parton distributions $\langle x^n \rangle_q(\mu)$, $\langle
x^n \rangle_{\Delta q}(\mu)$, and $\langle x^n \rangle_{\delta q}(\mu)$ are
scheme and scale dependent, and we will convert our lattice matrix elements to
the $\overline{MS}$ scheme and evaluate them at the scale  
$\mu^2 = \frac{1}{a^2} \sim $ 4 GeV$^2$.

\subsection{Lattice operators}

Our objective is to calculate matrix elements of traceless, appropriately
symmetrized and antisymmetrized operators of the general form
\begin{equation}
{\cal O} \equiv \bar{\psi}\Gamma \gamma_{\mu_1}
{\stackrel{\,\leftrightarrow}{D}}_{\mu_2}\cdots 
{\stackrel{\,\leftrightarrow}{D}}_{\mu_n}\psi ,
\end{equation}
where $\Gamma = 1, \gamma_5$, or $\gamma_5 \gamma_{\sigma}$,  on a
hypercubic lattice to approximate the corresponding continuum operators as
accurately as possible. Hence we choose  representations of the hypercubic
group, H(4) \cite{bib-mandula},  to eliminate operator mixing as much as possible,
and after fulfilling this objective, to minimize statistical errors by including as
few nonzero components of the nucleon momentum as possible.

Since H(4) is a subgroup of the Lorentz group, irreducible representations of the
Lorentz group  are in general reducible under the H(4)
group, and we choose the representation to optimize the
approximation. It is essential to choose a representation that does not mix with
lower dimension operators, since the coefficients would increase as $1/a^n$ in
the continuum limit. In addition,  because of the possible  inaccuracy of
perturbative mixing coefficients and the difficulty of determining mixing
coefficients nonperturbatively, it is desirable to avoid mixing with
operators of the same dimension as well. In choosing between operators
with the same mixing properties, it is desirable to use a nucleon source
with as few non-zero spatial momentum components as possible, since
each projection introduces substantial stochastic noise. In fact, we will
subsequently show that with available configurations, it is not possible to
obtain adequate statistics in any momentum sector other than $\vec p = 0$.
Since any expectation value of an operator with tensor index $j$ is
proportional to 
$P_j$ (or $S_j$ for spin dependent quark distributions), the nucleon must 
have an additional momentum component projection for each new distinct
tensor index that is added to an operator.   Hence, the goal is
to limit the number of distinct spatial indices. Eventually, as one proceeds
to higher moments of quark distributions, all the space-time indices are
exhausted and it becomes impossible to avoid mixing with lower dimension
operators.

The representations we have chosen for our operators using these criteria are
enumerated in Table \ref{tab-operators}. To illustrate the selection process, we
describe selection of the spin-independent operators and analogous analysis
yields the  remaining operators.

To measure $\langle x \rangle_q $ one needs to calculate 
matrix elements of the traceless part of the operator 
$\bar{q} \gamma_{\{\mu} {\stackrel{\,\leftrightarrow}{D}}_{\nu\}} {q}$,
which belongs to the representation (1,1) in the continuum decomposition
$$
(\frac{1}{2},\frac{1}{2})\otimes(\frac{1}{2},\frac{1}{2})=(0,0)
\oplus(1,0)\oplus(0,1)\oplus(1,1).
$$
On the lattice, the nine dimensional representation (1,1) splits into two 
irreducible representations, {\bf 3}$_1^+$ and {\bf 6}$_3^+$, both of which 
are symmetric and traceless, where the notation for representations is described
in the caption of Table \ref{tab-operators}.  As a consistency check, it is
desirable to calculate operators from each representation.
For  the first operator, denoted $\langle x \rangle_q^{(a)} $, we select  
the basis vector of {\bf 6}$_3^+$:
$$
\langle P|\bar{q} \gamma_{\{1} {\stackrel{\,\leftrightarrow}{D}}_{4\}} 
{q}|P\rangle = 2\langle x \rangle_q^{(a)}  \cdot P_{\{1} P_{4\}} ,
$$
and for $\langle x\rangle_q^{(b)} $, we choose the basis vector of 
{\bf 3}$_1^+$:
$$
\langle P|\bar{q} \gamma_{4} {\stackrel{\,\leftrightarrow}{D}}_{4} {q} 
- \frac{1}{3}(\bar{q} \gamma_{1} {\stackrel{\,\leftrightarrow}{D}}_{1} {q}
+\bar{q} \gamma_{2} {\stackrel{\,\leftrightarrow}{D}}_{2} {q}+\bar{q} 
\gamma_{3} {\stackrel{\,\leftrightarrow}{D}}_{3} {q})|P\rangle 
= 2\langle x\rangle_q^{(b)} \cdot (P_{4} P_{4}-\frac{1}{3}\vec{P}^2) .
$$
Note that since $\langle x\rangle_q^{(b)} $ involves $\gamma_{4}
{\stackrel{\,\leftrightarrow}{D}}_{4}$, it can be measured with $\vec P = 0$
whereas since $\langle x\rangle_q^{(a)} $ involves $\gamma_{1}
{\stackrel{\,\leftrightarrow}{D}}_{4}$, it requires a state projected onto non-zero
$P_1$.

For $\langle x^2 \rangle_q $, none of the three (symmetric) 
representations {\bf 4}$_1^-$ is appropriate, since they are not traceless 
and hence mix with lower-dimensional operators. The only representations with
two distinct indices are the one {\bf 8}$_1^+$, which is not symmetric and must
therefore be rejected, and the two {\bf 8}$_1^-$ 's

$$
\bar{q}\left(\gamma_{4}{\stackrel{\,\leftrightarrow}{D}}_{1}
{\stackrel{\,\leftrightarrow}{D}}_{1} - \frac{1}{2}(\gamma_{4}
{\stackrel{\,\leftrightarrow}{D}}_{2}
{\stackrel{\,\leftrightarrow}{D}}_{2}+\gamma_{4}
{\stackrel{\,\leftrightarrow}{D}}_{3}
{\stackrel{\,\leftrightarrow}{D}}_{3})\right){q} ,
$$
and 
$$
\bar{q}\left(\gamma_{1}{\stackrel{\,\leftrightarrow}{D}}_{4}
{\stackrel{\,\leftrightarrow}{D}}_{1}+\gamma_{1}
{\stackrel{\,\leftrightarrow}{D}}_{1}
{\stackrel{\,\leftrightarrow}{D}}_{4}
-\frac{1}{2}(\gamma_{2}{\stackrel{\,\leftrightarrow}{D}}_{4}
{\stackrel{\,\leftrightarrow}{D}}_{2}+\gamma_{2}
{\stackrel{\,\leftrightarrow}{D}}_{2}
{\stackrel{\,\leftrightarrow}{D}}_{4}+\gamma_{3}
{\stackrel{\,\leftrightarrow}{D}}_{4}
{\stackrel{\,\leftrightarrow}{D}}_{3}+\gamma_{3}
{\stackrel{\,\leftrightarrow}{D}}_{3}
{\stackrel{\,\leftrightarrow}{D}}_{4})\right){q}.
$$
which mix as discussed in 
Refs.~\cite{bib-bianchi,bib-qcdsf-renorm,bib-qcdsf-h4}.

For $\langle x^3 \rangle_q $ the following representations have positive charge
conjugation and do  
not mix with  lower dimensional operators:
{\bf 2}$_1^+$, {\bf 2}$_2^+$, {\bf 3}$_2^+$, {\bf 3}$_2^+$, {\bf 3}$_3^+$,  
{\bf 6}$_2^+$ and {\bf 6}$_4^+$.
However, the only representations that require a single non-zero momentum
component are the two {\bf 2}$_1^+$'s, which generically could mix with each
other but do not mix at the one loop level for Wilson or overlap
fermions~\cite{bib-qcdsf-renorm,Capitani:2000aq,Capitani:2000xi}.

Note that in addition to the mixing discussed above, in full QCD there is also
mixing between gluonic operators and flavor-singlet fermion operators 
for moments of the quark density and helicity that will
not be considered in this work because we have not yet evaluated lattice
matrix elements of the relevant gluon operators.

\section{Perturbative Renormalization}

Since the phenomenological light cone quark distributions with which we
compare our lattice results are extracted from experimental data using  the
$\overline{MS}$  renormalization scheme, we have converted our lattice
calculations  to the 
$\overline{MS}$ scheme in 1-loop perturbation theory using
\begin{equation}
O^{\overline{MS}}_i(Q^2)=\sum_j\left(\delta_{ij}+\frac{g_0^2}{16\pi^2}\,
\frac{N_c^2-1}{2N_c}
\left(\gamma^{\overline{MS}}_{ij}\log(Q^2a^2)-(B^{LATT}_{ij}
-B^{\overline{MS}}_{ij})\right)\right)\cdot O^{LATT}_j(a^2) .
\label{app-renorm1}
\end{equation}
The anomalous dimensions $\gamma_{ij}$ and the finite constants $B_{ij}$
are given in Table \ref{tab-renorm1} for Wilson fermions and the specific
operators  considered. 

The $Z$ factors that convert lattice results  to the 
$\overline{MS}$ scheme at scale $Q^2 = 1/a^2$ are equal to
\begin{equation}
Z(g_0^2=6/\beta)=1-\frac{g_0^2}{16\pi^2}\,\frac{4}{3}
\left(B^{LATT}-B^{{\overline{MS}}}\right) ,
\end{equation}
and are tabulated for two typical values of  $\beta$. 
Note that all of the moments of quark distributions calculated in this work will
be presented at the scale of $\mu^2 =$ 4 GeV$^2$ in the $\overline{MS}$ scheme.

Details of perturbative renormalization may by found in
Refs.~\cite{bib-Kronfeld,Capitani:2000wi,bib-qcdsf-renorm}.   The results in
Table
\ref{tab-renorm1} are taken from Refs.~\cite{Capitani:2000wi,Capitani:2000aq},
in which the renormalization factors of $\langle x \rangle _{\delta q} $
and  $d_1$ for Wilson fermions were calculated for the first time and the
remaining operators were checked with earlier results in
Refs.~
\cite{Capitani:2000xi,bib-cap0,bib-bianchi,bib-qcdsf-main,%
bib-qcdsf-renorm,Brower:1996pe},
revealing a discrepancy in the case of
$\langle x^3
\rangle_q$.

\section{Lattice Calculations}

In this section, we describe salient aspects of our  lattice  calculation of
proton matrix elements of the operators corresponding to moments of parton
distributions. Details may be found in Ref.~\cite{bib-dolgov-thesis}.

\subsection{Connected Diagrams}

\begin{figure}[tp]

$$
\BoxedEPSF{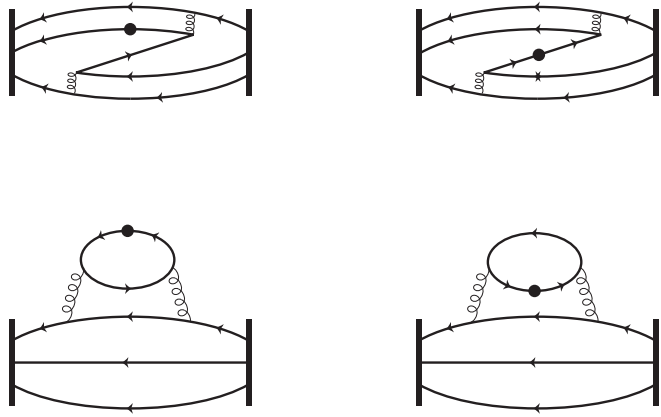 scaled 900}
$$
\vspace*{.2cm }
\caption{Connected (upper row) and disconnected (lower row) diagrams contributing
to hadron matrix elements. The left column shows typical contributions of quarks and
the right column shows contributions of antiquarks.}
\label{HadrMatrElem}  
\vspace*{.2cm} 
\end{figure}

Proton matrix elements of the operators in Eq.~\ref{ourmoments} are
calculated by evaluating the connected and disconnected diagrams shown in
Fig.~\ref{HadrMatrElem}. Note that the term "disconnected" refers to
diagrams in which the quarks are disconnected but of course as shown in 
Fig.~\ref{HadrMatrElem}, the overall Feynman diagram is still connected by
gluons. It is important to recognize
at the outset that although the connected diagrams involve three propagators
naively corresponding to the "valence" quarks in a simple hadron model, the
contributions of the connected diagrams do not necessarily correspond to the
valence parton distribution $q_{val} \equiv q -\bar{q} $ defined in
phenomenological analyses of high energy scattering data.  Rather, as shown in
Eq.~\ref{mom1}, the odd moments of $\langle x^n \rangle_q$ and $\langle
x^n
\rangle_{\delta q}$ and even moments of $\langle x^n \rangle_{\Delta q}$
measured in deep inelastic lepton scattering
correspond to the sum $ q +\bar{q} $ and only the even moments of $\langle x^n
\rangle_q$ and $\langle x^n
\rangle_{\delta q}$ and odd moments of $\langle x^n \rangle_{\Delta q}$
correspond to the difference $ q -\bar{q} $.

It is technically much more
difficult to evaluate the disconnected diagrams than connected diagrams and
results using the  eigenmode expansion technique that has recently been
developed  for this purpose  \cite{Neff:2001zr} are not yet available. Therefore,
in the present work, we will compare the contributions of connected diagrams with
flavor non-singlet combinations of experimental results. Because the
coupling of the disconnected loop to the rest of the diagram is flavor
independent,  the disconnected diagrams do not contribute to the difference
between the moments for degenerate up and down quarks.  For example,  the axial
charge can be calculated directly from differences of connected contributions,
$g_A = \langle
\Delta u  + \Delta \bar{u} -
\Delta d - \Delta \bar{d} \rangle _{connected} $. Hence, 
we will
subsequently compare our  results with moments of flavor non-singlet
combinations of the sum or differences of phenomenological quark and antiquark
distributions.

One technical difference between our calculations and  those of other works is
the fact that we use Dirichlet boundary conditions for the valence fermions in
the time direction. With periodic or antiperiodic boundary conditions,
contributions from the images of the sources and sinks  restrict the useful
source-sink separation to less than half the total time extent.  Dirichlet boundary
conditions prevent propagation from these image sources and allow utilization of
a larger fraction of the lattice volume for ground state operator measurements. 
We note that the closely related Schr\"odinger functional boundary
conditions used in a calculation of the average quark momentum in the pion
provide similar benefits~\cite{Guagnelli:2000em}.

\subsection{Sources}
\label{lattice_calculation_sources}
To facilitate measurement of proton matrix elements, we have
optimized the overlap between a computationally efficient nucleon source and
the proton ground state. The starting point is the  interpolating field for
the proton 
\begin{equation}
J =\epsilon^{abc}\,\left[u^a\,C\gamma_5\,d^b\right]\,u^c , 
\label{app-nucl1}
\end{equation}
where $C$ denotes the charge conjugation matrix. This field  corresponds to the 
non-relativistic quark model wave function  
$\epsilon^{abc}\,\left[u^a_{\uparrow}d^b_{\downarrow}-u^a_\downarrow 
d^b_\uparrow\right]\,u^c_\uparrow$ in the non-relativistic
limit.  Since truncation of the lower components
does not significantly reduce its overlap with the physical proton
\cite{bib-grandy-thesis} we save an overall factor of two in computation and
storage by only calculating propagators from the upper components of the source. 

As shown below in Section~\ref{lattice_results}, if $J$ in Eq.~\ref{app-nucl1} is a
point source, the overlap with the nucleon ground state is of the order
$10^{-4}$, so generalization to finite spatial extent is desirable. Since the source
must be factorizable in order for matrix elements to be calculated from single
quark propagators, we use distributed quark fields with adjustable spatial
extent. In ref
\cite{bib-pochinsky-thesis}, it was shown that gauge fixed sources with gaussian
wave functions and gauge invariant smeared Wuppertal sources 
\cite{bib-wupsm1,bib-wupsm2} produced comparable results, and for
computational convenience we use the latter.

\begin{figure}[ht]
$$
\BoxedEPSF{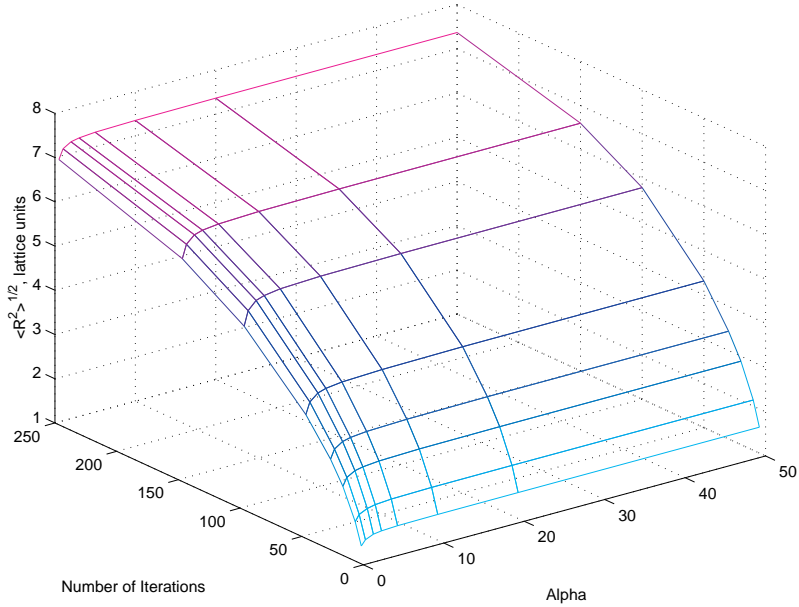 scaled 600}  
$$
\caption{The rms radius of a gauge invariant smeared quark source as a
function of the coefficient $\alpha$ and number of smearing steps $N$ defined in
Eq.~\ref{opt-psi1Hpsi}. }
\label{smeared_source}
\end{figure}

The smeared wave function is
defined \cite{Frommer:1995ik}

\begin{equation}
\psi^{(N,\alpha)} = (1+\alpha H)^N\psi^0
\text{\hspace{0.15in}where\hspace{0.15in}}
H = \sum_{i=1}^3(U_{n, i}+U^\dagger_{n-\hat{i}, i})
\text{\hspace{0.15in}and\hspace{0.15in}}
\psi^0(n)=\delta_{nn_0}
\label{opt-psi1Hpsi}
\end{equation}
with the spatial extent being controlled by the coefficient of the nearest
neighbor hopping term, $\alpha$, and the number of smearing steps, $N$.  
The distribution is approximately Gaussian with an equilibrated gauge
field $U$ producing a narrower distribution than for the free case. A convenient
measure of the smearing is the  rms radius 
$$  r_{rms} =\langle
r^2\rangle^{\frac{1}{2}}=\left[\frac{\int\,d^3x\,r^2\,
\psi^*\psi}{\int\,d^3x\,\psi^*\psi}\right]^\frac{1}{2}, \label{rmsrad}
$$
 and figure
\ref{smeared_source} shows how  $r_{rms}$ depends on the parameters
$N$ and $\alpha$. As one expects from the free case and from the fact that
smearing is a random walk governed by the gauge fields, the rms radius is
approximately proportional to $\sqrt{N}$. Note that the size of the source is
nearly independent of $\alpha$ for $\alpha > 3$, at which point the constant
term in Eq.~\ref{opt-psi1Hpsi} becomes negligible relative to the hopping
term. In the calculations described in section~\ref{lattice_results}, we set
$\alpha = 3$ and use $N$ to adjust $r_{rms}$ to optimize the source.

The source is optimized by maximizing the overlap between the normalized state
created by the action of the source on the QCD vacuum $|\Psi_J\rangle =
{\cal N} \bar{J} |\Omega
\rangle$ and the normalized ground state of the proton $| 0 \rangle$.  Denoting
the momentum projected normalized eigenstates of the proton by $| n \rangle $
and their energies by $E_n$, the momentum projected two point correlation
function may be expanded:
\begin{equation}
\langle J(t) \bar{J}(0) \rangle = {\cal C} \sum_n \left|\langle 
\psi_J|n\rangle\right|^2 e^{-E_nt}\label{twoptfn}
\end{equation}
where ${\cal C}$ is an unknown normalization constant. Since one can directly
measure the correlation function at zero time separation
\begin{equation}
A = \langle J(0)\bar{J}(0)\rangle =  {\cal C} \sum_n \left|\langle J
|n\rangle\right|^2 \label{Acoef}
\end{equation}
and reliably fit the large t behavior of the correlation function to extract the
ground state contribution
\begin{equation}
B =  {\cal C}\left|\langle J |0\rangle\right|^2, \label{Bcoef}
\end{equation}
the probability that the source contains the proton
ground state is given by 
\begin{equation}
P(0) = \frac{B}{A} =\left|\langle \Psi_J |0\rangle\right|^2.\label{P(0)}
\end{equation}

The importance of maximizing the overlap of the source with the proton is
particularly clear when one considers the three point function used to calculate
matrix elements:
\begin{equation}
\langle J(t_3) {\cal O}(t_2) J(t_1) \rangle = {\cal C} \sum_{n,m}\langle 
\psi_J| n \rangle \langle n | {\cal O } | m \rangle \langle m | \psi_J \rangle 
e^{-E_n (t_3 - t_2) -E_m (t_2 - t_1)}
\label{three_pt_fn}.
\end{equation}

Although the contribution of a contaminant $\langle \psi_J | n \rangle \sim
\epsilon$,  to the two point function, Eq.~\ref{twoptfn},
is of order
$\epsilon^2$, the contribution to the three point function, Eq.~\ref{three_pt_fn}
is of order
$\epsilon \frac{\langle n | {\cal O } | 0 \rangle}{\langle 0| {\cal O } | 0 \rangle}$.
Depending on the off-diagonal elements of any particular operator, the error
when the operator is close to one of the sources may be substantial, and the sign
of the contribution is undetermined. Specific examples will be seen in
plateau plots presented in Section
\ref{lattice_results}.

\subsection{Sequential Propagators} 

There are two alternative strategies for calculating the connected diagrams in
Fig.~\ref{HadrMatrElem} with sequential propagators. Using a propagator from a
fixed source to all points $x_0$ times the operator ${\cal O}(x_0)$ as a new
source to propagate to an arbitrary sink location allows one to calculate
matrix elements of the operator between all source sink separations
\cite{bib-liu,Gusken:1999xy,bib-fuji,bib-aoki,%
bib-sesam-sigma,Maiani:1987by}.
The matrix element is then obtained from the linear term in the time
separation between the source and sink.  The SESAM collaboration has used
this variable time extent method to calculate the axial charge
\cite{Gusken:1999xy} for the configurations used in this work, and we will
compare our results with theirs in Section~\ref{lattice_results}.  However,
the fact that a new set of propagators must be calculated for each operator
makes this alternative too costly for the large set of operators of interest in
this present work. 

The alternative we will use is to combine two forward propagators from a
fixed source, smear them and momentum project on a specific time slice to
create a sink, and use this sink as a new source for a backward going
propagator \cite{bib-bernard}. The matrix element for any operator can then be
obtained by combining forward and backward propagators with that operator.
Because the time separation between source and sink is fixed in this approach,
the effect of this separation has been studied and optimized as described in
Section~\ref{lattice_results}. The details of calculating matrix elements with
these sequential propagators are described in Appendix A.

\subsection{Lattice Data Analysis}
\label{lattice_calculations_data_analysis}
To ensure that the systematic and statistical errors are well understood and
controlled, we have used several methods to analyze the lattice measurements
of matrix elements.  

In our primary analysis,  which is used for all the results presented in
tables and graphs, we define a central window for the location of the
insertion of the measured operator
$[t_{source}+\delta t, \,t_{sink}-\delta t]$ in which contamination from excited
states in the source or sink has negligible effect, average the ratio of
three-point to two-point functions over this window, and use the
jackknife method \cite{bib-jkn} to estimate the average value and variance. 
For smeared sources, we used
$\delta t=3$ and for tests with unsmeared sources we have also used
$\delta t=5$.

As a secondary analysis, to ensure that we have avoided systematic errors
from excited state contaminants in the plateau region, we have also fitted the
phenomenological form
\begin{equation}
R(t)=R_0+\sum_i\left(b_ie^{-c_it}+b_ie^{-c_i(T-t)}\right)
\label{opt-fit}
\end{equation}
over the whole range between the source and the sink. 
The exponentials correcting for the contaminants are symmetric because we
have used the same smearing in the source and sink, and the value for the matrix
element is $R_0$.

From our data, it was only possible to determine one pair of 
exponentials, and we used a jackknife estimate on the fitted values of $R_0$
to determine the error.  For operators for which the errors were small enough
that the exponential contaminants were  well determined and stable,
the plateau and exponential fits produced statistically consistent results with
comparable errors. This ensures that our primary results do not have
statistically significant bias from excited states. In some cases, the exponential
contaminants were poorly determined and produced spurious fluctuations  in
the analysis which overestimated the actual statistical fluctuations in the plateau
region. 
This is the reason we selected the plateau fits as our primary
analysis
and view the exponential fits as a secondary consistency check.

\begin{figure}[ht]
$$
\BoxedEPSF{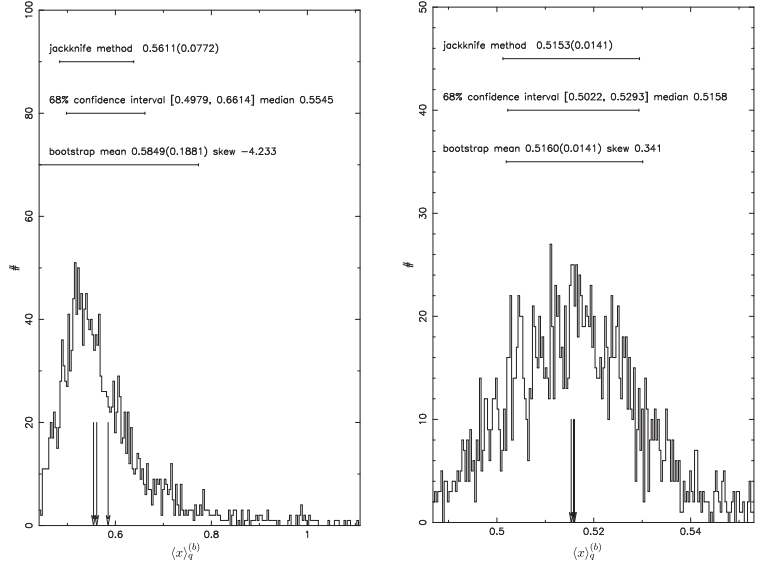 scaled 800}  
$$
\caption{Bootstrap distributions of $\langle x \rangle _q^{(b)}$ for  ensembles of
25  (left plot) and 204 (right plot) SESAM configurations at $\kappa=0.1575$}
\label{bootstrap}
\end{figure}

To further ensure that we understood our statistical errors, we also examined
bootstrap distributions  \cite{bib-recipes} for all our observables at the lightest
SESAM quark mass. An ensemble of  size $M $ was created by randomly
picking configurations from the original sample with replication, and typical
distributions of values of measurements of $ \langle x \rangle _q^{(b)}$ 
are shown in Fig.~\ref{bootstrap}, where $M = 2000$ and the
sample sizes in the left and right figures are  25 and  204
configurations  respectively.
Three error bars are shown in each case: the basic jackknife error bar  which is
calculated independently of the bootstrap procedure, the 68\%  confidence
interval based on the bootstrap distribution, and the statistical  variance of
the bootstrap measurements. The three arrows correspond to the jackknife
result, the median of the distribution, and the mean of the distribution.  This
figure shows that for a sample size as small as  25, the distribution is far from
Gaussian, the three error bars differ significantly, and even the arrows disagree
slightly. For the large sample size 204 used in our full data analysis, everything is
consistent: the distribution is nearly Gaussian, the three errors are consistent, and
the three arrows are equivalent. Thus, the bootstrap analysis gives added assurance
that the statistics are understood and under control.

\section{Lattice Results}
\label{lattice_results}

This section presents the results of our calculations of moments of
quark distribution functions in full and quenched QCD, as well as relevant tests
of the lattice technology and consistency checks.

\subsection{Lattice Gluon Configurations}

The parameters specifying the full QCD,  quenched, and cooled configurations 
used in our calculations are tabulated in Table \ref{t_datasets}. The full QCD
configurations were calculated using the hybrid Monte Carlo algorithm by the
SESAM collaboration\cite{bib-sesam-spectro}   at  $\beta = 5.6$ and by the SCRI
collaboration\cite{bib-edwards-spectro} at $\beta = 5.5$. The quenched
configurations were calculated at MIT at $\beta = 6.0 $ to be directly
comparable in lattice spacing to the unquenched $\beta = 5.6$ configurations,
and the cooled configurations were obtained by applying 50 cooling sweeps to
the SESAM configurations as described later.

The practical limitations associated with these configurations pose significant
problems in extrapolating to both the chiral and continuum limits as required to  
compare with phenomenological data.

\leftline{\bf Chiral extrapolation}

A common criterion to keep the pion correlation length significantly smaller
than the physical size of the lattice and to avoid unphysical interactions with
periodic images is to require that the pion Compton wavelength be less than
one-fourth the spatial dimension of the lattice.
Because resources limit us to $16^3 \times 32$ lattices, at $\beta = 5.6$ where
the lattice spacing $a = 0.091$ fm, the lattice dimension is only 1.46 fm so 
that the pion mass must be greater than 540 MeV. However even for the lowest
SESAM quark mass, which exceeds this criterion, there are significant
discrepancies between mass measurements on  $16^3 \times 32$ and $24^3
\times 48$ lattices\cite{bib-orth}, indicating serious finite volume effects. Hence,
in our final analysis, we have only included the heaviest three SESAM quark
masses in our results. 

As discussed in detail in the final section, much of the physics of the pion cloud
of the nucleon is omitted on a lattice of size 1.46 fm with pions heavier than 540
MeV, so it is unreasonable to expect naive linear extrapolation of calculations in
this regime to accurately include the quantitatively important physics of the
pion cloud.  However, since we do not presently have data in a regime in which
we can fully determine the parameters of an extrapolation incorporating the
behavior known from chiral perturbation theory, most of our data analysis
will be based on naive linear extrapolation. This extrapolation is still useful
because it crisply frames the issues that need to be confronted in comparison
of {\it ab initio} lattice calculations with experiment.

\leftline{\bf Continuum extrapolation}

Due to limitations in presently available configurations for Wilson fermions,
extrapolation to the continuum limit is even more problematic. 
Table \ref{t_world-dyn-wil} summarizes parameters and lattice
spacings for configurations that are presently available for dynamical Wilson
fermions.  Since we are calculating nucleon properties, we believe it is most
consistent to set the scale using the lattice spacing determined from the nucleon
mass, $a_N$. Unfortunately, there is a large discrepancy in determinations of 
$a_N$, and for qualitative purposes we have assumed the behavior given by the
LANL results, since this is the only data set in which one group calculated
consistently at three different $\beta$.

\begin{figure}[!ht]
        \centering
\BoxedEPSF{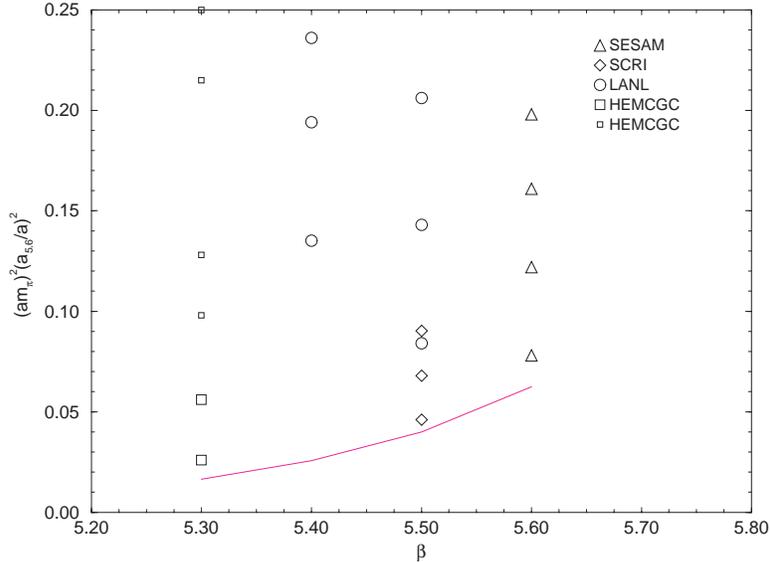 scaled 470}
        \caption{ The scaled pion mass squared,
$(a m_\pi)^2 \times (a_{5.6}/a)^2$, for the dynamical Wilson fermion
calculations tabulated in Table \ref{t_world-dyn-wil}.  The dotted line
shows the mass at which the pion Compton wavelength equals one-fourth of
the spatial dimension.}
        \label{fig-world-dyn-wil-mq_vs_beta}
\end{figure}

One goal has  been to use a comparison of calculations of moments of quark
distributions using SCRI configurations at $\beta$ = 5.5 and SESAM
configurations at $\beta$ = 5.6  to obtain an indication of the finite lattice size
dependence, and we report results  from both $\beta$'s below. However, the
problem of finite size effects is shown in Fig.~\ref{fig-world-dyn-wil-mq_vs_beta},
where we present scaled values of $m_\pi^2$ at different values of $\beta$ for
the data sets in Table \ref{t_world-dyn-wil}. The scaled mass squared,
$(a m_\pi)^2 \times (a_{5.6}/a)^2$, converts the mass in different lattice units at
each
$a$ to fixed lattice units at $\beta = 5.6$ for ease of comparison. 
Thus, the ordinate is proportional to $m_{\pi}^2$ in physical units, and 0.05
corresponds to $m_{\pi}^2 = 0.24$ GeV$^2$ or  $m_{\pi} = 490$ MeV.  The
dotted line shows the point at which the pion Compton wavelength equals one
fourth of the spatial dimension.  Given that the lowest SESAM point has significant
finite size errors, we expect the lowest two SCRI points to have comparable
contamination from finite size effects. Since there are no SCRI points comparable to
the highest SESAM points, there is no way to delete the lowest SCRI points and
perform an extrapolation comparable to the SESAM extrapolation.  Hence, we
believe that the differences between moments of quark distributions calculated
with the SCRI and SESAM configurations is an undetermined combination of
lattice spacing errors and finite volume effects. New dynamical Wilson quark
calculations are underway at $\beta = 5.5$ and $\beta = 5.3$ to separate
these finite volume and finite lattice spacing effects.

\subsection{Source optimization}

To obtain as much physics as possible from the SESAM and SCRI
configurations, the nucleon sources have been optimized in their overlap with
the ground state and with respect to their separation in Euclidean time.

\begin{figure}[tp]

$$
\BoxedEPSF{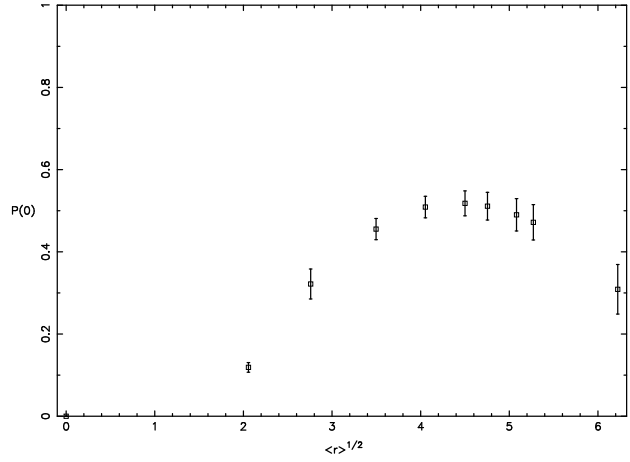 scaled 470}
$$
\vspace*{0pc}
\caption{Overlap between  a smeared source and the proton ground state as
a  function of
the source rms radius. The overlap for zero smearing is  $6 \times10^{-5}$.}
\label{overlap}  
\vspace*{0cm} 
\end{figure}

As discussed in section \ref{lattice_calculation_sources}, the rms radius, 
$\langle r^2
\rangle^{\frac{1}{2}}$, characterizing the spatial extent of a Wuppertal source
was varied to maximize the overlap of the source with the ground state
proton wave function, 
$P(0) = \left|\langle \Psi_J |0\rangle\right|^2$.
Figure \ref{overlap} shows the overlap, with jackknife errors,  between the smeared
source and the proton ground state as a function of the source rms radius in
lattice units,  calculated at $\beta = 5.6 $, $\kappa = 0.1575$,
and smearing parameter $\alpha = 3$. 
The number of smearing steps, $N$, to produce the rms radii in this plot
ranges from 20 at the lowest non-zero point to 250 for the highest.  It is
interesting that the overlap increases from $6 \times 10^{-5}$ for a point
source to 0.5 for an rms radius of the order of 4.5 lattice units. Clearly,
this four-order-of-magnitude increase is a dramatic aid in improving the
plateau for measurements. Based on these and analogous calculations at 
$\beta = 5.5$, the optimal Wuppertal smearing occurs at a physical rms radius
of 0.4 fm, which corresponds to $\alpha$= 3 and $N = 50$ at $\beta = 5.6$
and  $\alpha$= 3 and $N = 30$ at $\beta = 5.5$.

Note that this overlap calculation
opens the possibility of performing instructive variational studies of the
nucleon ground state.  Already, a very simple quark and gluon wave
function made up a product of three gaussian single-particle wave functions
smeared with surrounding links of glue has a   50\% overlap with the full
ground state and is measured with good statistical accuracy. Hence, it
should be straightforward and practical to investigate variationally the
extent to which the wave function can be improved by the addition of
changes in the quark wave function, such as the addition of diquark
correlations, or of changes in the gluon wave function.

The optimal separation between source and sink is a compromise between
two competing effects. As the separation increases, the size of the usable
plateau 
increases, reducing systematic error and in principle allowing for increased
statistical accuracy from averaging over an increasingly large region.
However, the propagators themselves are exponentially decreasing with
separation and the fractional error with which they can be measured
increases exponentially. (This is analogous to the exponential error growth in
the measurement of Wilson loops or Polyakov lines, but because the
measurement of the three-point function is intrinsically non-local, it is not
amenable to the  error reduction techniques of L\"uscher and Weisz
\cite{Luscher:2001up}).

\begin{figure}[ht]
$$
\BoxedEPSF{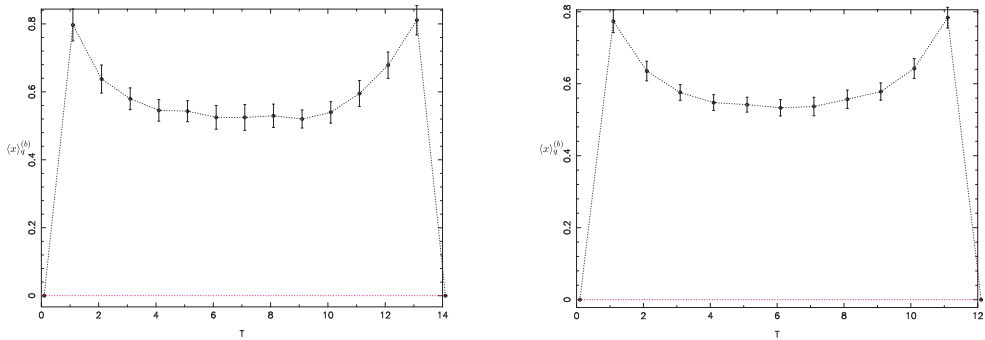 scaled 500}  
$$
\caption{Plateaus for $\langle x \rangle_q^{(b)}$ calculated with point sources
separated by 14 (left) and 12 (right) time steps using 100 configurations at
$\beta = 5.6$ and $\kappa = .1575$. }
\label{plateau}
\end{figure}

To determine the optimal compromise between these two competing effects
more quantitatively,  plateaus were calculated as a function of the
source-sink separation for point sources, and the results for $\langle x
\rangle_q^{(b)}$ using 100 configurations at
$\beta = 5.6$ and $\kappa = .1575$ are shown in Fig.~\ref{plateau} for time
separations $\Delta T = 12$ and 14. The errors for $\Delta T = 14$ are 50\% larger than for 
$\Delta T = 12$, and become prohibitive for larger separations.   
This is consistent with the results of ref~\cite{bib-qcdsf-main}, that for a
$16^3\times32$ lattice in quenched QCD with 
$\beta=6.0$, which has a comparable lattice spacing,  the error bars grow
significantly for
$\Delta T>13$. Although the
plateau region for $\Delta T = 14$ is hardly present at $\Delta T = 12$ for point sources,
one may still extract the correct matrix element using the exponential fit
method described in section \ref{lattice_calculations_data_analysis}.
Furthermore, using optimal smeared sources to enhance the ground state
overlap by four orders of magnitude decreases the excited state
contaminants to the point that a well defined central plateau is recognized for
$\Delta T = 12$ as shown below. Hence, to reduce the overall statistical errors in our
calculations, it is optimal to use a physical separation of  approximately 1.1 fm
which corresponds to $\Delta T=12$ for SESAM configurations at  $\beta=5.6$
and to
$\Delta T=10$ for SCRI configurations at  $\beta=5.5$.

\begin{figure}[ht]
$$
\BoxedEPSF{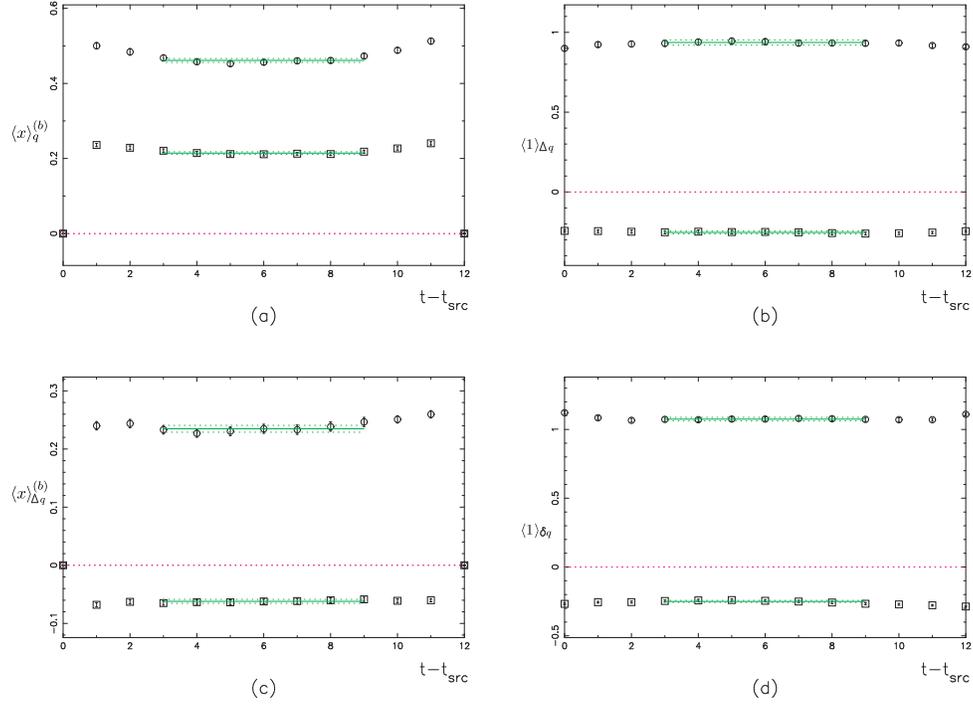 scaled 500}  
$$
\caption{ Plateaus obtained using optimally smeared sources with $\alpha =
3 $ and $N = 40 $ separated by 12 time steps on SESAM configurations with
$\beta=5.6$ and
$\kappa = 0.1560$.
Measurements as a function of Euclidean time of the
operators $\langle x \rangle_q^{(b)}, \langle 1 \rangle_{\Delta q}, 
\langle x \rangle_{\Delta q}^{(b)}$, 
and $  \langle 1 \rangle_{\delta q} $ are
shown in panels a, b, c, and d respectively. Circles and squares denote matrix
elements for up and down quarks respectively, the error bars are smaller than the
symbols and shown within them, and the solid lines denote fits within the plateau
region.}
\label{SESAM_plateaus}
\end{figure}

The final quality of the plateaus used for measurements of operators in a
zero momentum ground state are shown in Fig.~\ref{SESAM_plateaus}, where
we plot measurements  of the
operators $\langle x \rangle_q^{(b)}, \langle 1 \rangle_{\Delta q}, 
\langle x \rangle_{\Delta q}^{(b)}$, 
and $  \langle 1 \rangle_{\delta q} $ as a function of Euclidean time for an
ensemble of 200 SESAM configurations with $\beta=5.6$ and
$\kappa = 0.1560$.  The improvement in the plateau for $\langle x
\rangle_q^{(b)}$ produced by smearing is seen by comparing the right panel
of Fig.~\ref{plateau} and the upper left panel of  Fig.~\ref{SESAM_plateaus}.
Note also, as emphasized in connection with Eq.~\ref{three_pt_fn}, that the
sign of the exponential contaminant near the source or sink may be either
positive or negative, with a small negative contribution arising in the case of
the axial charge, $\langle 1 \rangle_{\Delta q}$ shown in 
the upper right panel of Fig.~\ref{SESAM_plateaus}. These well-defined
plateaus are typical of all our measurements of operators in a
zero momentum ground state and show why we obtain consistent results
with plateau and exponential fits and satisfactory statistics. In the case of
projection onto non-zero momentum ground states, errors are substantially
larger. Although we also report these non-zero momentum results in this
work, the errors are generally unsatisfactory and additional calculations will
be carried out to improve their statistics.

One other optimization was investigated, but turned out to produce minimal
improvement. There is considerable freedom in selecting lattice operators
from the irreducible representations tabulated in Table \ref{tab-operators}.
For example, instead of the basis vector $\bar{q} \gamma_{4} 
{\stackrel{\,\leftrightarrow}{D}}_{4} {q} - \frac{1}{3}(\bar{q} \gamma_{1} 
{\stackrel{\,\leftrightarrow}{D}}_{1} {q}
+\bar{q} \gamma_{2} {\stackrel{\,\leftrightarrow}{D}}_{2} {q}+\bar{q} 
\gamma_{3} {\stackrel{\,\leftrightarrow}{D}}_{3} {q})$ of representation  {\bf
3}$_1^+$, one could equally well chose 
$\bar{q} \gamma_{4} 
{\stackrel{\,\leftrightarrow}{D}}_{4} {q} -\bar{q} \gamma_{i} 
{\stackrel{\,\leftrightarrow}{D}}_{i} {q}$ for $i = $ 1, 2, or 3. Since the
direction
$i = 3$ is inequivalent to the other two spatial directions because the spin is
polarized in the 3 direction, one can find numerically the optimal
combination of the basis vectors that minimizes the variance in the
measurement of $\langle x \rangle _q^{(b)}$. However, in the end, the
reduction in the statistical error is only about 7\%, so this refinement was not
implemented in our production calculations.

\subsection{Consistency checks}

This section summarizes a number of consistency checks that have been
performed to ensure the reliability and accuracy of our calculations of
moments of quark distributions.

\leftline{\bf Sources}

\begin{figure}[ht]
$$
\BoxedEPSF{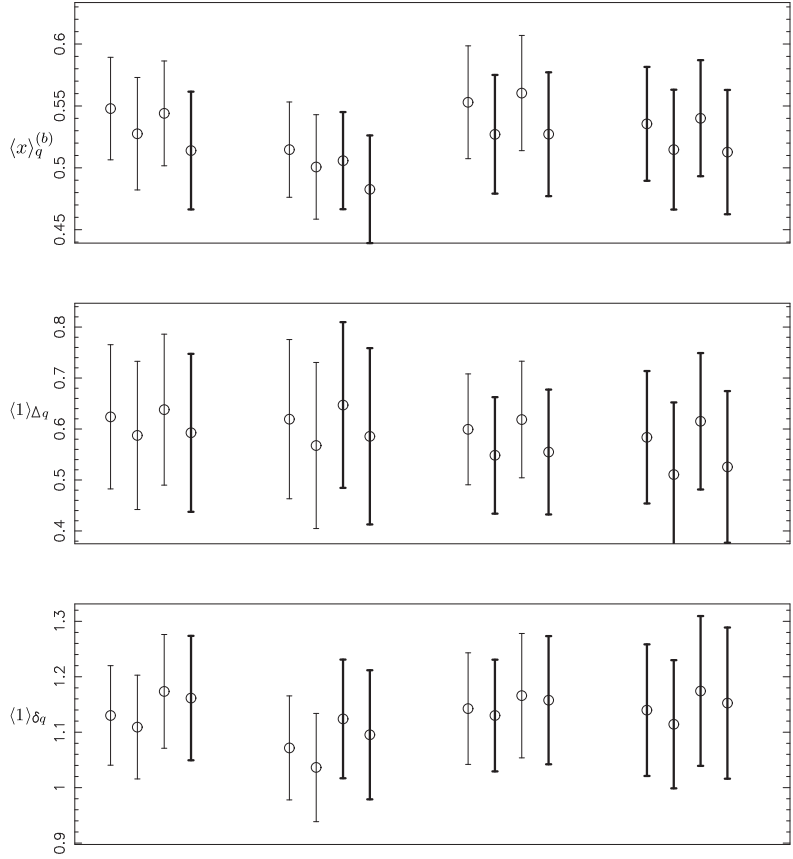 scaled 600}  
$$
\caption{Comparison of operator measurements using combinations of point
and smeared sources and sinks. From left to right, the four clusters, each
containing four error bars, correspond to point-point, point-smeared,
smeared-point  and smeared-smeared source-sink combinations respectively.
Within each cluster, from left to right,  the four error bars correspond to
windows in which the number of lattice points omitted from the window at
the source and sink are (3,3), (3,5), (5,3), and (5,5) respectively. All smeared
sources and sinks have $ N = 20 $  and $ \alpha =3 $. From top to bottom, the
panels show measurements of the operators $\langle x \rangle_q^{(b)}, 
\langle 1 \rangle_{\Delta q}$, 
and $  \langle 1 \rangle_{\delta q} $.}
\label{point_smeared_sources}
\end{figure}

One potential pitfall of optimizing sources to increase the size of the plateau
region is the possibility that although excited state contaminants appear to
have been diminished, unobserved systematic errors may have sneaked into
the plateau measurements.  To ensure that such systematic errors have not
been introduced into our present calculations, we have evaluated 
the operators of interest for various combinations of source and sink
combinations.  Figure \ref{point_smeared_sources} compares the results for 
$\langle x \rangle_q^{(b)},  \langle 1 \rangle_{\Delta q}$, 
and $  \langle 1 \rangle_{\delta q}$ calculated using  point-point,
point-smeared, smeared-point  and smeared-smeared source sink
combinations, showing that in all cases the results are statistically consistent.
Forty configurations at $\beta = 5.6$ were used with a source-sink separation
of 12, and the smeared sources had
$N = 20 $ smearing steps. Each cluster of four error bars corresponds to four
different window sizes over which the measurements were averaged,
characterized by the number of points omitted at the source and sink.
The case shown in Fig. ~\ref {SESAM_plateaus}, denoted (3,3),  corresponds to
omission of 3 points at the source and 3 points at the sink and is shown
at the left of each cluster. With a separation of 12 lattice spacings, there are
13 points in all, of which 6 are deleted leaving a measurement plateau of 7.
The remaining error bars in each cluster correspond to (3,5), (5,3), and (5,5)
points omitted at the source and sink respectively.  Thin lines denote 
cases in which the window comes too close
to either a point sink or point source and overlaps with
a region with contamination from excited states and the error bars denote
jackknife errors.

\begin{figure}[ht]
$$
\BoxedEPSF{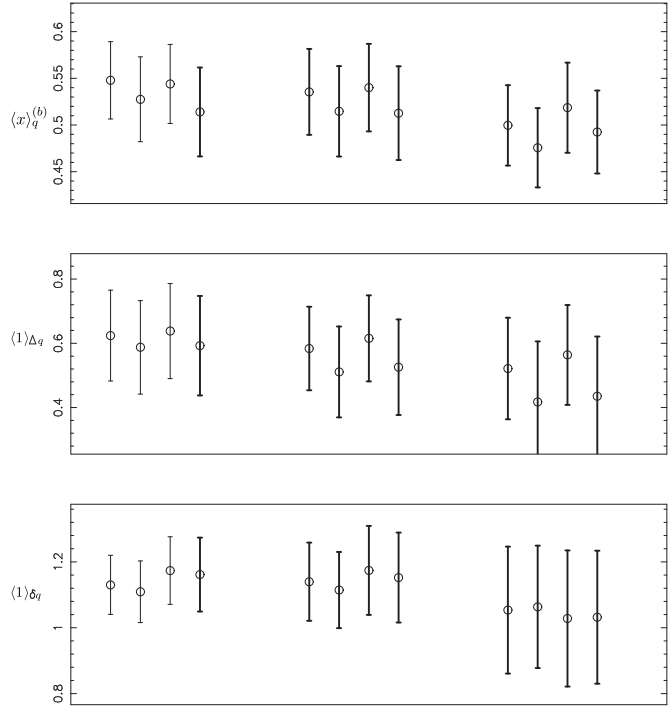 scaled 500}  
$$
\caption{Comparison of operator measurements using smeared sources of
different sizes. From left to right, the three clusters, each containing four
error bars, correspond to  $ N = 0$, $ N = 20$, and  $ N = 100$.  The
windows within each cluster and the operators are the same as in
Fig~\ref{point_smeared_sources}.          }
\label{N_smears}
\end{figure}

Figure \ref{N_smears} shows a related comparison of the 
measurements of the same three observables in which the source and sink
are smeared with $ N = 0$, $ N = 20$, and  $ N = 100$ smearing steps. The
window sizes within each cluster of error bars are the same as in the
previous case. Again, the observables are consistent for all three cases,
although the error bars become substantially larger for the largest smearing
because the size of the paths of link variables generated by the smearing
increases significantly.

\leftline{\bf Boundary conditions}

\begin{figure}[ht]
$$
\BoxedEPSF{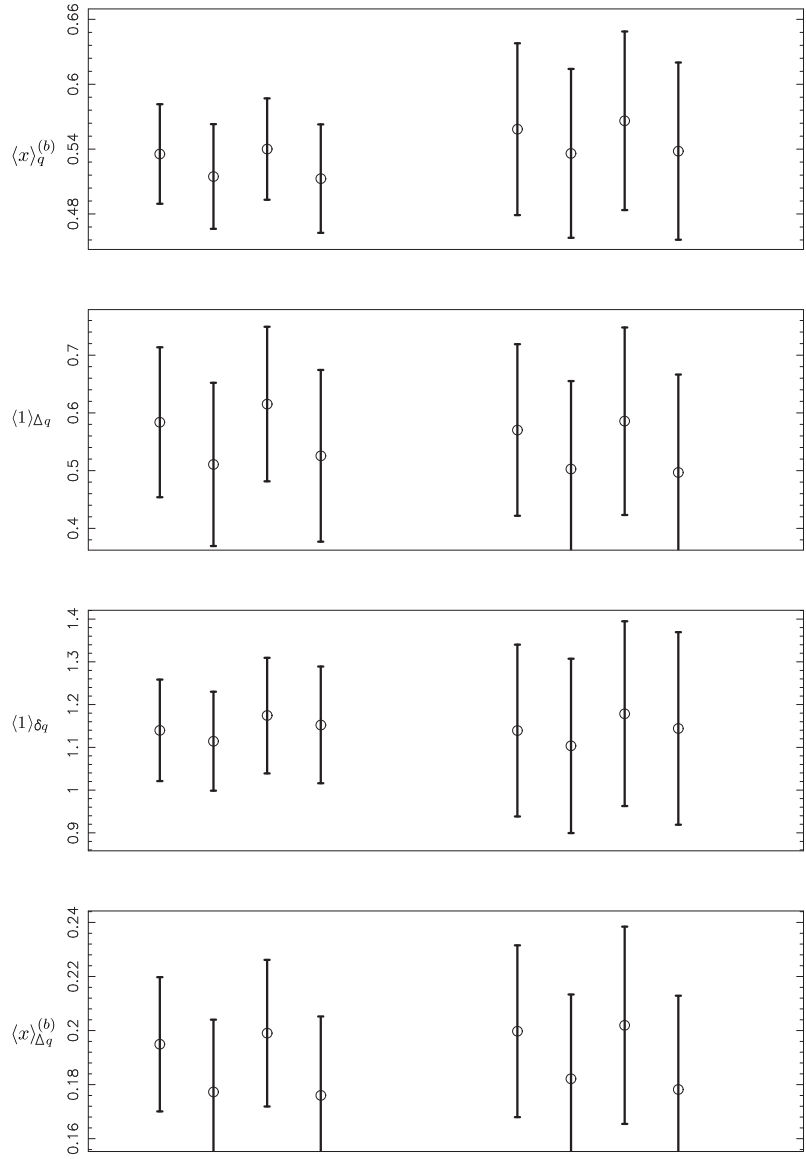 scaled 600}  
$$
\caption{Comparison of operator measurements using Dirichlet boundary
conditions (left cluster) and periodic boundary conditions (right cluster).  The
windows within each cluster and the operators for the top three panels are
the same as in Fig~\ref{point_smeared_sources}, and the bottom panel
corresponds to the operator $\langle x \rangle_{\Delta q}$. }
\label{boundary_conditions}
\end{figure}

The Dirichlet boundary conditions in Euclidean time we have used enable us
to calculate the exponential decay of two point functions far beyond the
midpoint of the lattice without the usual contributions of the propagation of
the parity partner of the nucleon from first images in the time direction. Only
within a few lattice points of the edge of the lattice do artifacts associated
with reflection from the boundary (or equivalently, negative image charges)
become significant. To verify that the choice of boundary condition does not
affect the measured values of hadronic matrix elements, 
Fig.~\ref{boundary_conditions} shows the results of measuring the operators 
$\langle x \rangle_q^{(b)},  \langle 1 \rangle_{\Delta q}, \langle 1
\rangle_{\delta q}$, and $\langle x \rangle_{\Delta q}$ using Dirichlet and
periodic boundary conditions.  Note that although the results are statistically
consistent, our choice of  Dirichlet boundary conditions significantly reduces
the error bars for some operators, such as $\langle x \rangle_q^{(b)}$ and 
$ \langle 1 \rangle_{\Delta q}, \langle 1 \rangle_{\delta q}$.

\leftline{\bf Precision}

\begin{figure}[ht]
$$
\BoxedEPSF{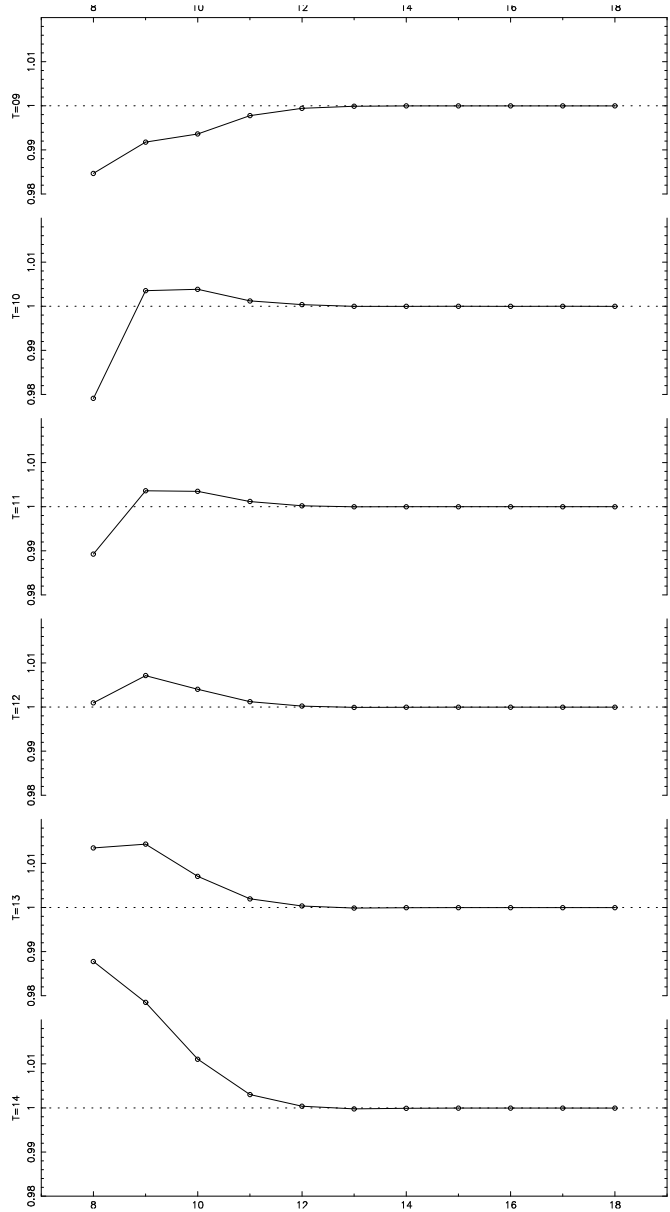 scaled 600}  
$$
\caption{Discrepancy between the axial charge, $\langle 1 \rangle_{\Delta q}$,
calculated in single precision (solid curve) and double precision (dotted
curve) as a function of $(-\log_{10}r^2_{min})$ where $r^2_{min}$ is the
conjugate gradient stopping residue. From the top panel to the bottom panel,
the source-sink separation increases from 9 to 14 lattice spacings. }
\label{precision}
\end{figure}

In our production calculations, the conjugate gradient inversion to
calculate quark propagators was performed in single precision (to increase
performance on the ES-40 cluster) using a perturbative stopping
residue $r^2_{min}= 10^{-14}$  and  four restarts\cite{bib-victor-thesis}. To
verify that single precision with the restart procedure provided adequate
precision, we show the discrepancy between calculations of the axial charge, 
$\langle 1 \rangle_{\Delta q}$,
calculated in single and double precision  as a function of the residual
 $r^2_{min}$ in Fig.~\ref{precision}.
We note that all discrepancies are already negligible at residue
$r^2_{min}= 10^{-12}$, providing a two order of magnitude safety margin.

\leftline{\bf Autocorrelation functions}

Autocorrelation functions were calculated for lattice measurements of 
$\langle x \rangle_q^{(b)}$  and $\langle x \rangle_{\Delta q}$
using the SESAM ensemble at $\kappa = 0.1575 $ consisting of 200
configurations  separated by 25 Hybrid Monte-Carlo
trajectories. Correlations for all non-zero configuration separations were
statistically consistent with zero, and all indications suggest that the 200
configurations are indeed independent for all the degrees of freedom
relevant for the present measurements.  

In contrast to the SESAM configurations, the SCRI configurations are only
separated by 10 trajectories. To minimize the possible effect of correlations,
the physical location of the source and sink was shifted by 16 time slices on
all odd configurations relative to those on even configurations. Autocorrelation
functions for 200 SCRI configurations at $\kappa = 0.1600 $ yielded a result
statistically inconsistent with zero only for separation by two configurations,
in which case the correlation was approximately $ 0.2 \pm 0.15$. Thus for
the degrees of freedom relevant to our proton matrix elements,
configurations separated by 10 Hybrid Monte-Carlo trajectories with sources
and sinks displaced by 16 time slices are independent, whereas those
separated by 20  Hybrid Monte-Carlo trajectories with the sources and sinks
at the same location still have non-vanishing correlations. The relative
independence of measurements at different source and sink positions
provides an attractive means of extracting additional physics from the
available full QCD configurations that will be exploited in subsequent
calculations.

\leftline{\bf Two and three point functions}

The nucleon two point function may be calculated in two ways: by combining 
at the sink three forward propagators calculated from the source, or by
combining a backward propagator with the  sequential source. The
agreement within roundoff error of these two results was used not only as a 
check on source methodology, but also at run time as a test to verify that stored
forward and backward propagators had not been corrupted.

Another check is to verify that the spatial integral  of the matrix element of
the conserved nonlocal vector current (the Noether current for the discrete
Wilson lattice action), ${\langle P|\int d^3x\,J^{(u), NL}_0|
P\rangle}/{\langle P| P\rangle}$, equals 2 or 1 for up and down quarks
respectively. This relation must be satisfied configuration by configuration
and is also used as a run-time test of the configurations.

An additional test of the source and sink construction is to project the sink
onto a fixed position in the sink time-slice, rather than momentum projecting
it, and to replace all link variables by their mirror images. We have verified
that the resulting three-point function is the mirror image of the original
result configuration by configuration.

Finally, as discussed previously, matrix elements of operators may either be
calculated with the variable time extent method, in which one sums over all
times at which the operator insertion could occur and determines the slope of
the dependence on $t_{sink}$, or by measuring the plateau using the fixed
source-sink separation as done in this work.  A consistency check that is
satisfied configuration by configuration  is that the sum
of the three-point function over the intermediate time at which the operator
is measured for a fixed source-sink separation should be equal to the
variable time extent  three point function for the fixed source-sink
separation. In addition, we have verified that our measurements of the axial
charge using the fixed source-sink separation method for the three largest
values of $\kappa$ agree quantitatively with those calculated by the
Wuppertal group using  the variable time extent method\cite{Gusken:1999xy}.

On the basis of these and other consistency checks, we are confident that the
calculations of moments of quark distributions reported in this work are both
reliable and accurate.

\subsection{Quenched results}

In order to investigate the effects of
quenching quantitatively, it is essential to compare quenched and unquenched
calculations at equal lattice spacing using precisely the same
computational methodology. Hence, we have calculated 
moments of polarized and unpolarized
parton distributions in quenched lattice QCD
at  $\beta=6.0$, which is comparable to full QCD at $\beta=5.6$.
These calculations also afford the
opportunity to check our results with the extensive calculations by the QCDSF
collaboration
\cite{bib-qcdsf-main,bib-qcdsf-tensor2,bib-qcdsf-a1,%
bib-qcdsf-spin,bib-qcdsf-d2,Gockeler:2000ja}. For Wilson fermions at
$\beta=6.0$, the critical 
$\kappa$ is  $\kappa_c=0.1572$, and we have performed our calculations at 
 $\kappa=0.1550,\,0.1540,\,0.1530$. Two of these values of  $\kappa$ coincide with
values used by  QCDSF, facilitating direct comparison of
unextrapolated as well as extrapolated results. In addition, the pion
masses corresponding to these
values of  $\kappa$  overlap the range of pion masses for the SESAM configurations,
providing a  meaningful comparison of quenched and unquenched
QCD.

\begin{figure}[ht]
$$
\BoxedEPSF{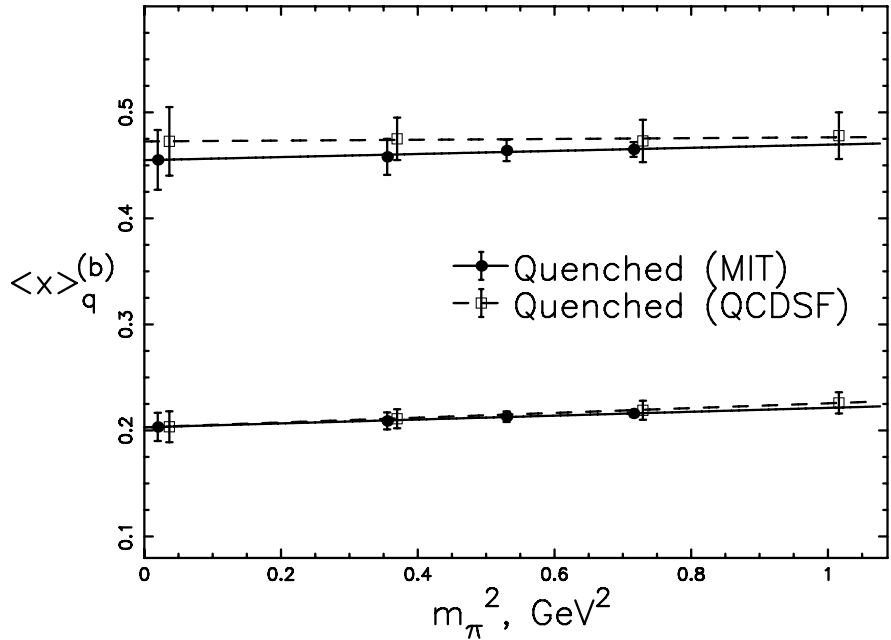 scaled 600}  
$$
\caption{Linear extrapolation of quenched calculations of the momentum
fraction
$\langle x \rangle _q^{(b)}$. The solid squares denote the results of the present
calculation, and  for comparison, QCDSF
results  are
shown by the open points.   Here and in subsequent figures, the upper and
lower curves correspond to up and down quarks respectively. }
\label{quenched_unpolarized_x}
\end{figure}

\begin{figure}[ht]
$$
\BoxedEPSF{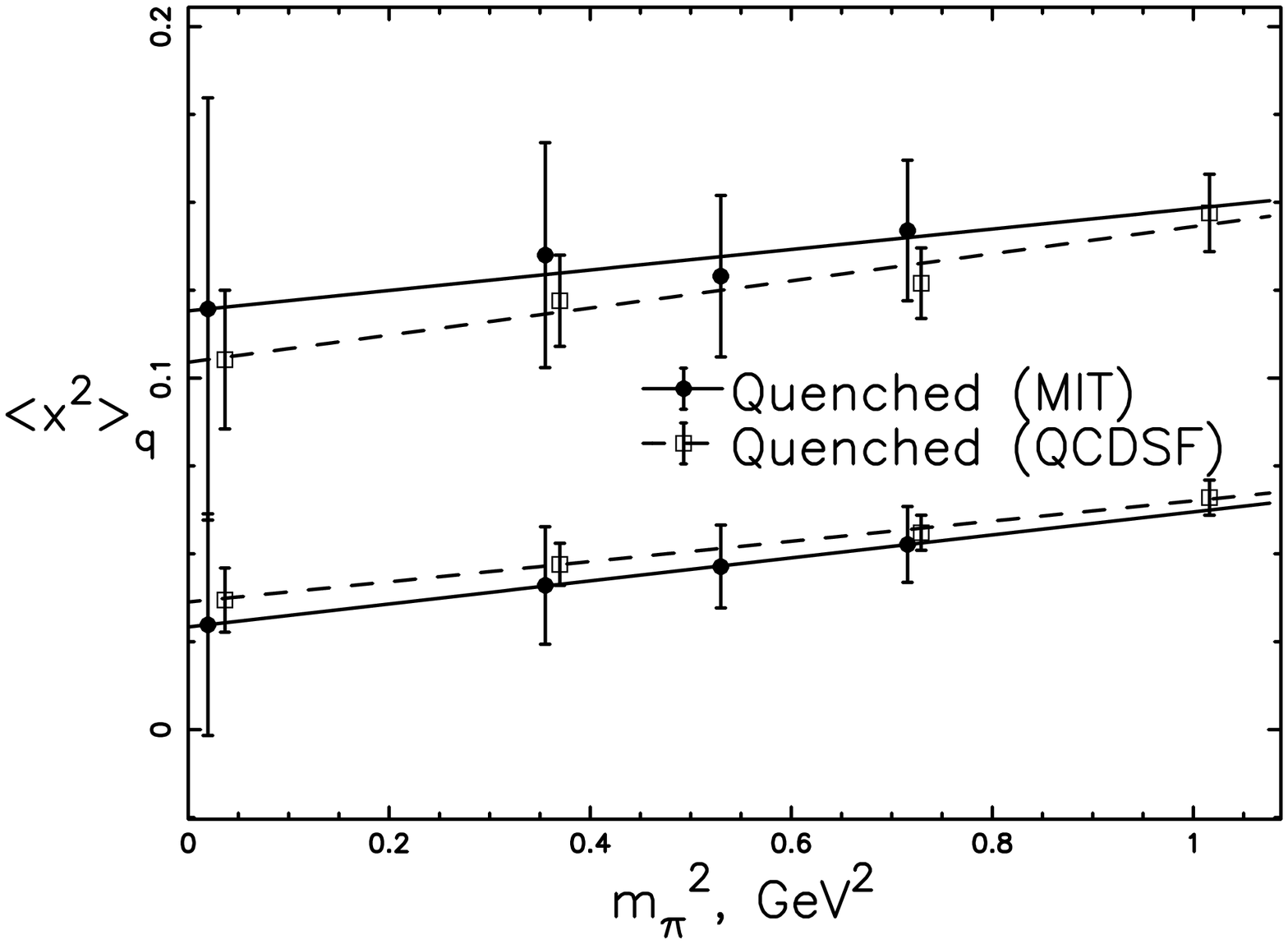 scaled 600}  
$$
\caption{Linear extrapolation of quenched calculations of
$\langle x ^2 \rangle_q$, where the symbols are as in
Fig.~\ref{quenched_unpolarized_x}.}
\label{quenched_unpolarized_xx}
\end{figure}

\begin{figure}[ht]
$$
\BoxedEPSF{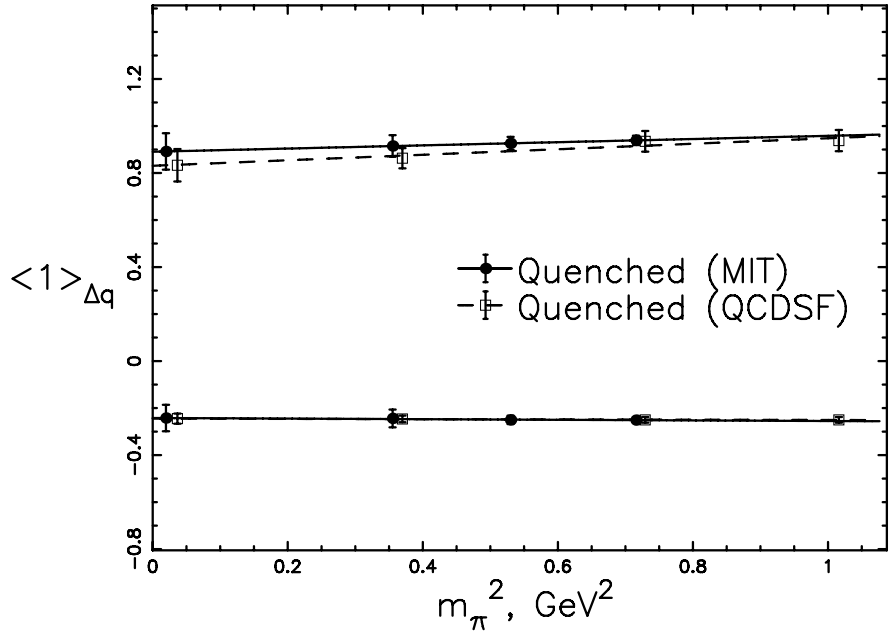 scaled 600}  
$$
\caption{Linear extrapolation of quenched calculations of the axial charge,
$\langle 1
\rangle _{\Delta q}$, where the symbols are as in
Fig.~\ref{quenched_unpolarized_x}.  }
\label{quenched_polarized}
\end{figure}

\begin{figure}[ht]
$$
\BoxedEPSF{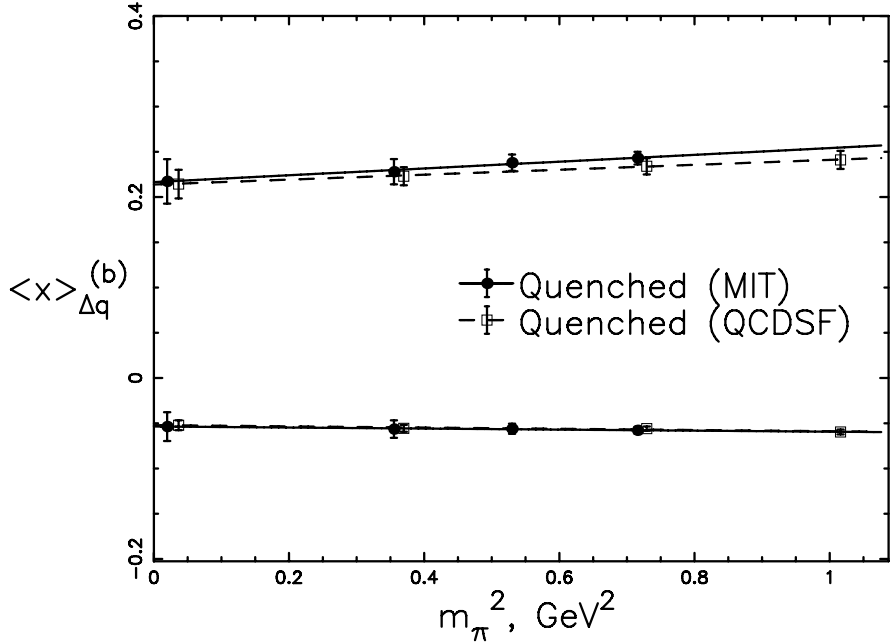 scaled 600}  
$$
\caption{Linear extrapolation of quenched calculations of the  moment
$\langle x
\rangle _{\Delta q}^{(b)}$, where the symbols are as in
Fig.~\ref{quenched_unpolarized_x}.  }
\label{quenched_polarized_x}
\end{figure}

Linear extrapolations of the quenched moments of
unpolarized up and down quark distributions
$\langle x \rangle _q^{(b)}$ and 
$\langle x ^2 \rangle_q$ are shown in Figs.~\ref{quenched_unpolarized_x} and
\ref{quenched_unpolarized_xx}.  Corresponding results for the quenched
moments of polarized quark distributions $\langle 1
\rangle _{\Delta q}$, and $\langle x \rangle _{\Delta
q}^{(b)}$ are shown in Figs.~\ref{quenched_polarized} and
\ref{quenched_polarized_x}. The complete numerical data for these and other
moments of quark distributions are tabulated in Table \ref{quenched_data}. All
results are presented in the $\overline{MS}$ scheme with $\mu^2= 4 \text{ GeV}^2$
and extrapolated linearly with a least squares fit to the chiral limit.  In these
figures, the measured values of
$m_{\pi}^2$ at each $\kappa$ were converted to physical units using $ \frac{1}{a} =
2$ GeV.

Our results for all operators agree within statistics with those of
QCDSF, and Figures \ref{quenched_unpolarized_x} --\ref{quenched_polarized_x}
show typical comparisons. 
Note that  for the three
operators that can be evaluated in a zero-momentum proton ground state, 
$\langle x \rangle _q^{(b)}$, $\langle 1
\rangle _{\Delta q}$, and $\langle x \rangle _{\Delta
q}^{(b)}$ the statistical errors in the extrapolated results are small and allow for
meaningful comparison with full QCD results and with phenomenology. 
In
contrast, the errors associated with the operator $\langle x ^2 \rangle_q$,
which requires projection onto a state with non-zero momentum, are so large
that meaningful comparison will require higher statistics.  Since QCDSF used
on the order of 1000 configurations whereas we only used 200
configurations, it is consistent that our error bars on individual points are
twice as large as theirs for this operator.  The fact  that the QCDSF
error bars are not half the size of ours for three matrix elements calculated  in a zero
momentum state as well is explained by the fact that the errors they reported for these
matrix elements are somewhat larger than jackknife errors because their original
analysis used a full error correlation matrix from which low eigenmodes were
removed \cite{horsleypc}.

\subsection{Full QCD Results}

\begin{figure}[ht]
$$
\BoxedEPSF{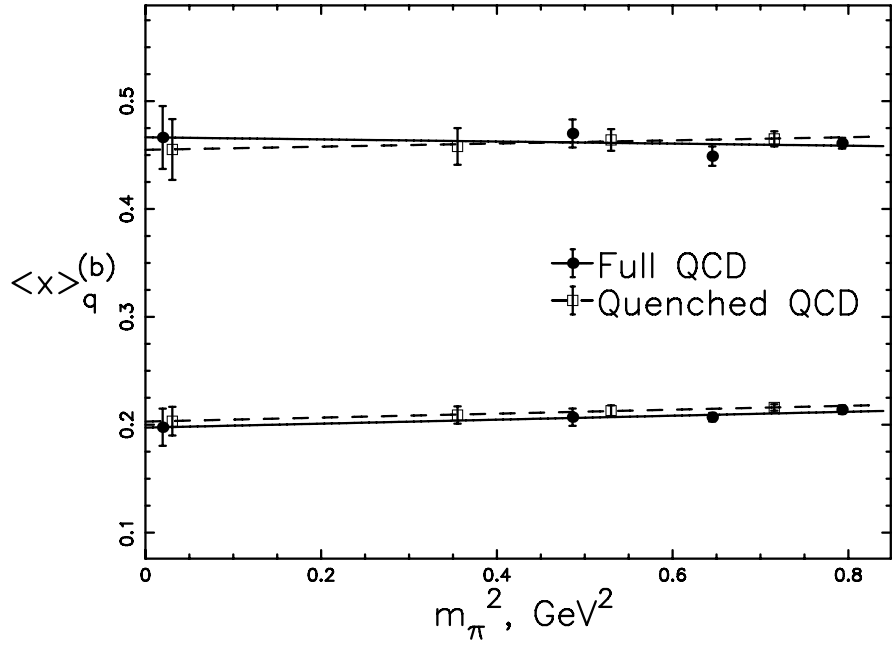 scaled 600}  
$$
\caption{Comparison of linear extrapolations of quenched and full QCD calculations  of
the momentum fraction, $\langle x  \rangle _q$.  The open symbols denote
quenched calculations at $\beta = 6.0$ and the closed symbols denote full QCD
calculations at
$\beta = 5.6$.  The upper and lower curves correspond to up and down quarks
respectively.}
\label{full_unpolarized_xq}
\end{figure}

\begin{figure}[ht]
$$
\BoxedEPSF{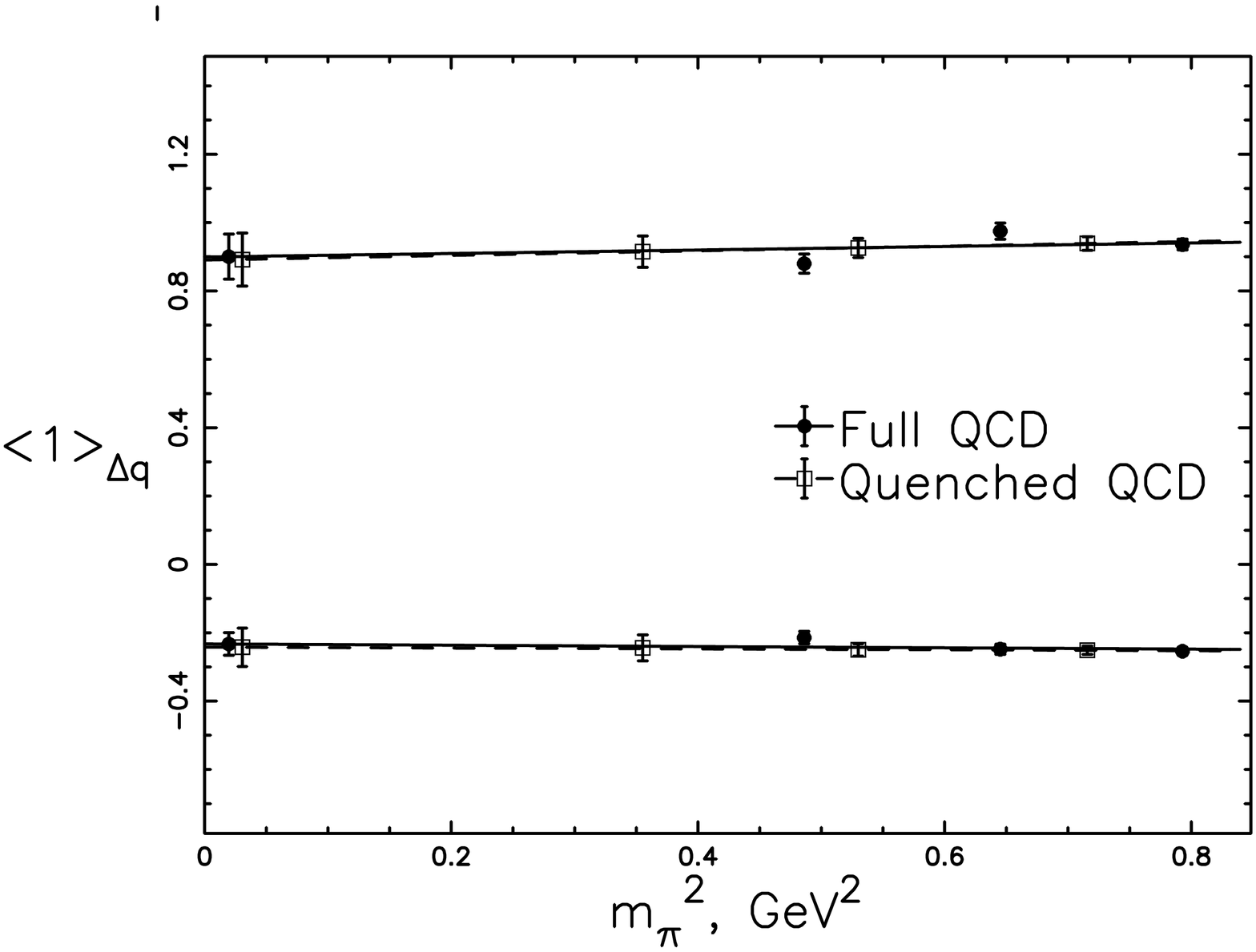 scaled 600}  
$$
\caption{Comparison of linear extrapolations of quenched and full QCD calculations  of
the axial charge, $\langle 1 \rangle _{\Delta q}$, where the symbols are as in
Fig.~\ref{full_unpolarized_xq}.}
\label{full_polarized_Delta_q}
\end{figure}

\begin{figure}[ht]
$$
\BoxedEPSF{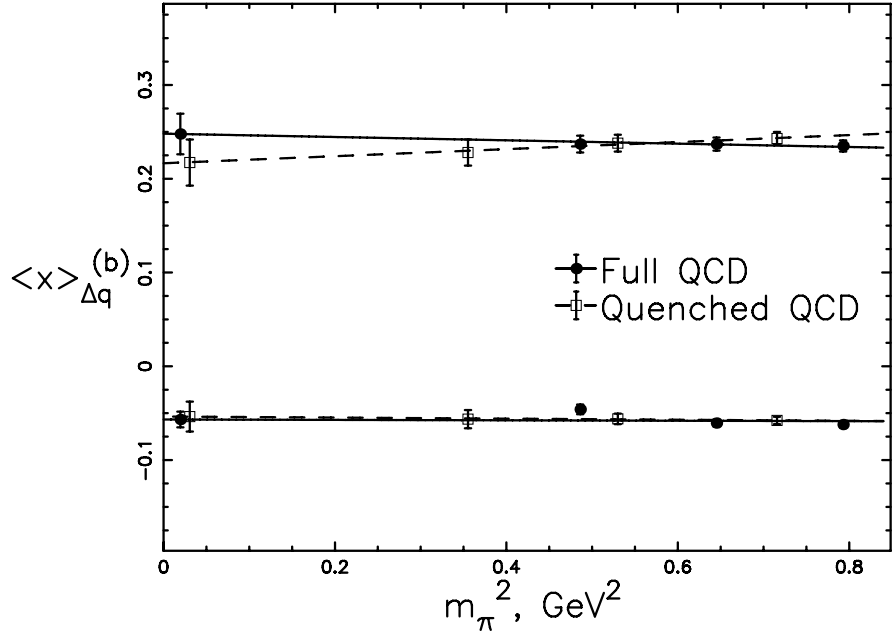 scaled 600} 
$$
\caption{Comparison of linear extrapolations of quenched and full QCD calculations  of
the first moment of the quark spin distribution, $\langle x \rangle _{\Delta q}^{(b)}$,
where the symbols are as in Fig.~\ref{full_unpolarized_xq}.}
\label{full_polarized_x_Delta_q}
\end{figure}

\begin{figure}[ht]
$$
\BoxedEPSF{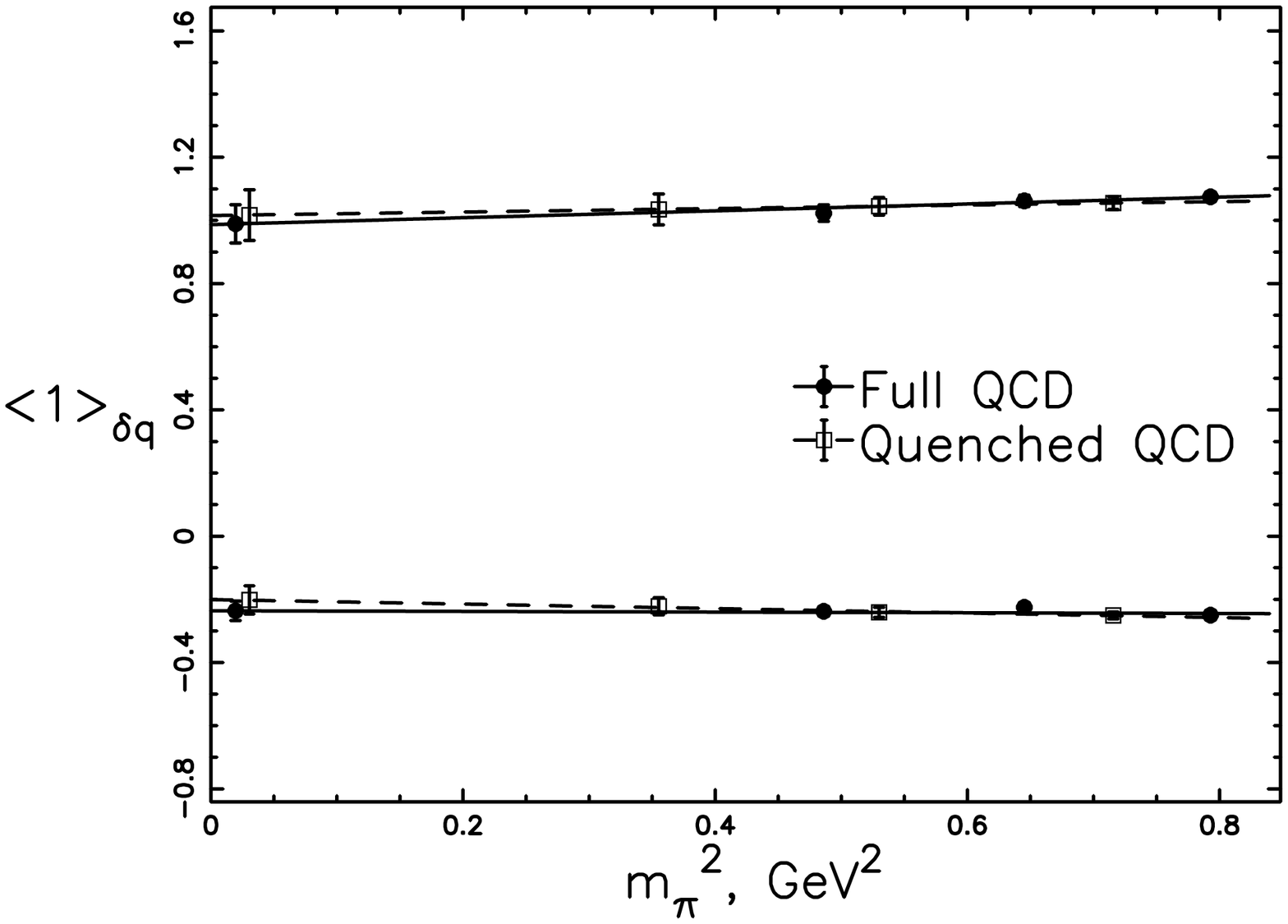 scaled 600}  
$$
\caption{Comparison of linear extrapolations of quenched and full QCD 
calculations  of the tensor charge,  $\langle 1 \rangle _{\delta q}$, where the symbols
are as in Fig.~\ref{full_unpolarized_xq}.}
\label{full_transversity_delta_q}
\end{figure}

A central result of this work is the fact that for the range of quark masses
accessible in the present calculations, quenched and full QCD calculations at
comparable lattice spacing agree within statistical errors. The moments of
quark distributions in full QCD at $\beta = 5.6$ calculated using SESAM
configurations are shown in Table~\ref{SESAM_data}. Note that to avoid finite 
volume effects, moments at the lightest  quark mass are not included in the
extrapolation.
Detailed comparison with 
Table~\ref{quenched_data} shows that in the case of all the operators we have
calculated, the full QCD results at $\beta = 5.6$ are statistically consistent with
those calculated in quenched QCD at $\beta = 6.0$. Furthermore, in the case of
operators that can be evaluated in zero-momentum nucleon states, the
statistical errors are sufficiently small that  any
differences between the linearly extrapolated quenched and unquenched
results are small compared with the discrepancies with experiment discussed
below.  The level of agreement between quenched and full QCD of $\langle x ^2 
\rangle _q$, $\langle 1 \rangle _{\Delta q}$, $\langle x \rangle
_{\Delta q}^{(b)}$, and $\langle 1 \rangle _{\delta q}$ is also shown in
Figs~\ref{full_unpolarized_xq},
\ref{full_polarized_Delta_q},
\ref{full_polarized_x_Delta_q}, and 
\ref{full_transversity_delta_q} respectively. Hence, the present work strongly
rules out the conjecture that the serious discrepancies between quenched
calculations and phenomenology arose from the omission of the effect of
dynamical quarks.  As noted previously, the magnitude of the errors in
observables measured in nucleons with non-zero momentum is qualitatively
larger, and no strong conclusions can be reached for these observables until
higher statistics calculations are carried out.

In addition to the calculations at $\beta = 5.6$, which correspond to the same
lattice spacing as our quenched calculations, we have also calculated the same
operators at 
$\beta = 5.5$ using the SCRI configurations in an attempt to study the approach
to the continuum limit. The results are tabulated in
Table~\ref{SCRI_data}. 
Note that by Eq.~\ref{app-renorm1}, the shifts in $a$ and $\beta$ in going 
from $\beta = 5.6$ to $\beta = 5.5$ produce changes in the renormalization
constants that are  negligible on the scale of the present statistical
errors, so these changes in renormalization factors have been omitted.
Unfortunately, as pointed out in connection with
Fig.~\ref{fig-world-dyn-wil-mq_vs_beta}, the two lightest quark masses lie in a
regime in which we expect significant errors due to finite-volume effects, so the
chiral extrapolations are physically suspect. Even so, the bulk of the operators
evaluated at $\beta = 5.5$  and  $\beta = 5.6$ agree within statistics. The only
exceptions, which are only slightly beyond one standard deviation, are $\langle
x ^2
\rangle _u$,
$\langle 1
\rangle _{\Delta u}$,
$\langle x \rangle _{\Delta u}^{(b)}$ , $\langle 1 \rangle _{\delta u}$, and
$\langle x \rangle _{\delta d}$. Because of the uncertainty in finite volume
effects and small lever arm, we have not attempted to extrapolate the present
data to the continuum limit.  Higher statistics calculations at smaller quark
masses for several coupling constants are presently underway to more seriously
address the behavior in the continuum limit.

\subsection{Cooled Results}

A fruitful strategy to obtain insight into hadronic physics is to use the numerical
evaluation of the sum over all quark and gluon configurations contributing to
the path integral to isolate those paths that dominate the action.  In recent
years, this approach has provided strong evidence that in QCD with light
quarks, topological excitations of the gluon field, which in the
semiclassical limit  correspond to instantons,
play a major role in hadron structure.  By minimizing the action
locally in a process known as cooling~\cite{bib-Berg},  the
instanton content of the quenched and full QCD vacuum has been
extracted~\cite{Negele:1998ev,Ivanenko:1997nb,Ringwald:1999ze}. 
Comparison of hadronic observables calculated with all gluons  and  those
obtained using only  the instantons remaining after cooling has demonstrated
qualitative agreement for  hadron masses, quark distributions, and vacuum
correlation functions of hadron currents~\cite{Chu:vi}. 
Calculation of the lowest quark eigenmodes has revealed zero modes
correlated spatially with the instantons and truncation of the
quark propagators to the zero mode zone has produced the full strength
of the $\rho$ and $\pi$ contributions to vacuum correlation functions
\cite {bib-ivan-thesis,Negele:1999mb}.

Hence, to obtain further insight into the structure of the
proton, we have used cooling to remove essentially all the gluon fluctuations
except for instantons from the full QCD configurations, and compared the quark
distributions calculated in full QCD and including only the contributions of
instantons. The role of instantons is particularly interesting in considering the
spin structure of the proton, since the  't Hooft instanton interaction  is the
only  vertex in QCD that directly removes helicity from valence
quarks and transfers it to gluons and quark-antiquark pairs and is therefore a
natural mechanism to explain the so-called ``spin crisis''
\cite{Forte:1990xb,Shore:1992pm,Kochelev:1997ux} .  
Our calculation of the contributions of
instantons to matrix elements of operators related by the operator
product expansion to deep inelastic scattering also closely parallels the
direct calculation of instanton contributions to deep inelastic scattering
in refs.
\cite{Ringwald:1994kr,Moch:1996bs,Ringwald:1998ek,Ringwald:2000gt}.
The present work
extends and substantiates earlier exploratory
investigations\cite{Dolgov:1998js}. 

To remove the essential non-instanton related gluon fluctuations without
producing unnecessary annihilation of instanton-antiinstanton pairs, we have
cooled the full QCD SESAM configurations at $\beta = 5.6$ using 50 cooling
steps.  This amount of cooling corresponds to roughly 25 cooling steps
for $\beta=5.7$,  which was an effective amount of cooling in ref~\cite{Chu:vi}.
Furthermore, calculations of nucleon two-point
functions using the SESAM configurations have shown that the results
with 25 and 50 cooling steps differ negligibly\cite{bib-victor-thesis}.

Because of the smoothness of cooled configurations,   small
statistical errors are obtained using an ensemble of 100 configurations, and
previous investigations have shown that the chiral dependence on quark mass
is quite linear in the region of interest. Hence, we cooled 100  configurations at 
$\kappa_{sea}=0.1570$ and at $\kappa_{sea}=0.1560$ for our comparison with
full QCD.  To make the cooled chiral extrapolation comparable to the uncooled
case, for each $\kappa_{sea}$, we selected the cooled valence quark mass,
$\kappa_{val}$, such that the ratio $\frac{m_\pi}{m_N}$ was the same as the
ratio in full QCD.
The resulting cooled values are $\kappa_{val}=0.1246$
for $\kappa_{sea}=0.1570$ , where $\frac{m_\pi}{m_N}=0.523$,  and
$\kappa_{val}=0.1235$ for $\kappa_{sea}=0.1560$,
where $\frac{m_\pi}{m_N}=0.476$,  and the full QCD masses were taken from
ref~\cite{bib-sesam-spectro}. Note that, because we are comparing two
theoretical calculations in the same physical volume, we have not discarded the
lightest quark mass case for full QCD out of concern for finite volume effects,
but rather have included it to enable comparison in the regime of the lightest
quark masses where dominance by the zero modes associated with instantons
should be most pronounced.

Since the high frequency quantum fluctuations are removed by cooling,  we
set all the  renormalization constants, $Z$,
to one in order to compare cooled results with the full QCD results. Although
we are aware of no rigorous argument as to  formulate renormalization in the
presence of cooling, this approximation appears the most physical and, for
example, is quite sensible for tadpoles, where we expect $U_0=\langle \frac{1}{3}
\text{Tr} U\rangle^{1/4}\simeq 1$.

\begin{figure}[ht]
$$
\BoxedEPSF{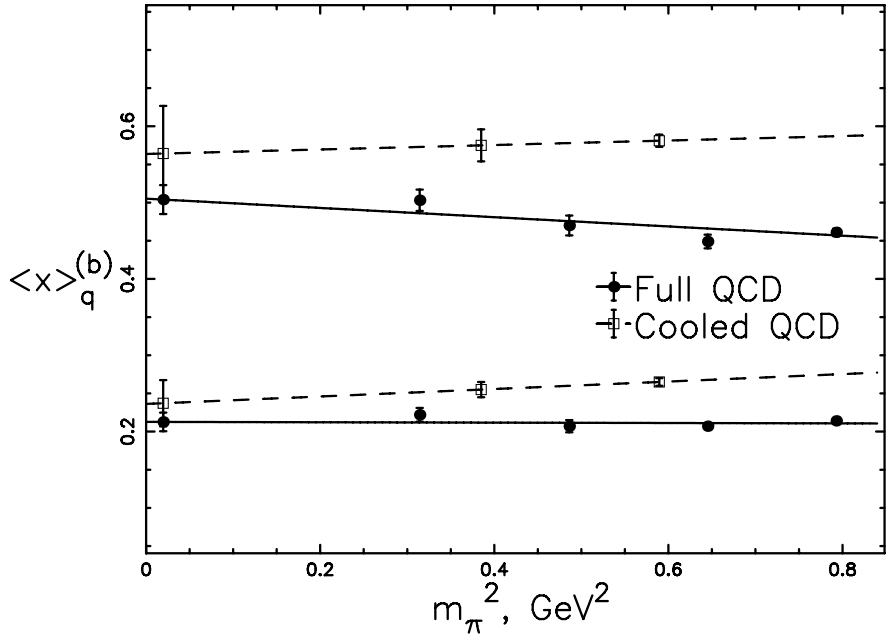 scaled 600} 
$$
\caption{Comparison of the quark momentum fraction $\langle x
\rangle _q^{(b)}$ calculated in full QCD and using configurations cooled to
eliminate essentially all contributions except those of instantons. Solid symbols
connected by solid curves denote full QCD and open symbols connected by
dashed curves denote results after cooling.  The upper and lower curves 
correspond to up and down quarks
respectively.}
\label{cooled_unpolarized_xq}
\end{figure}

\begin{figure}[ht]
$$
\BoxedEPSF{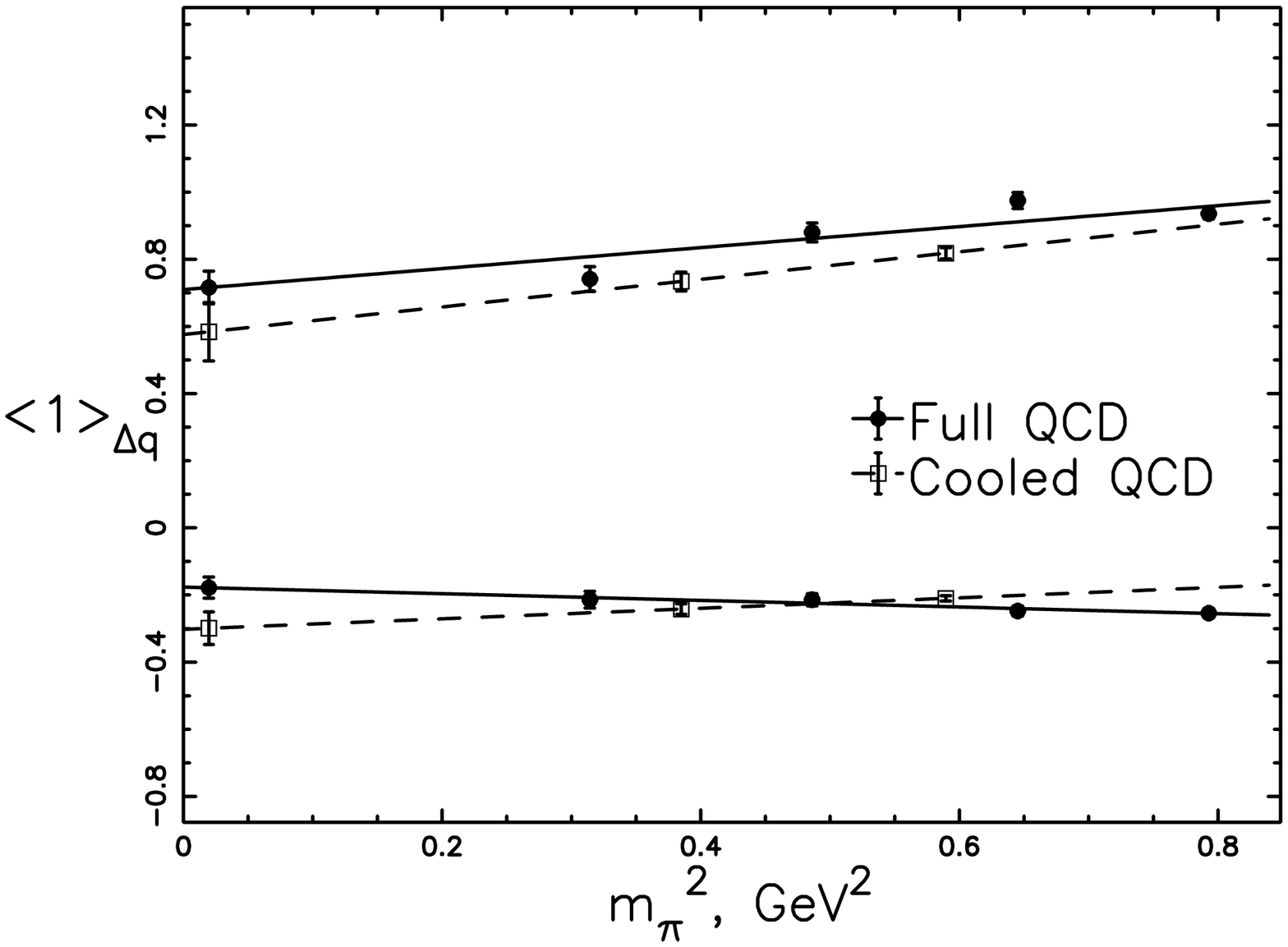 scaled 600}  
$$
\caption{Comparison of the axial charge $\langle 1
\rangle _{\Delta q}$   calculated in full QCD and using cooled configurations, as
in Fig.~\ref{cooled_unpolarized_xq}.}
\label{cooled_polarized_Delta_q}
\end{figure}

\begin{figure}[ht]
$$
\BoxedEPSF{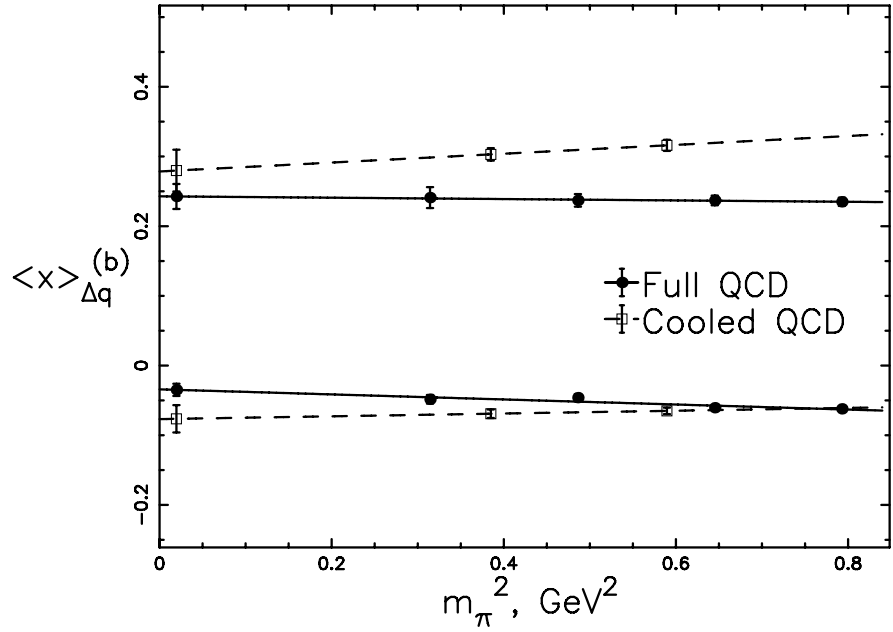 scaled 600} 
$$
\caption{Comparison of the first moment of the quark spin distribution, 
$\langle x \rangle _{\Delta q}^{(b)}$,   calculated in full QCD and using cooled
configurations, as in Fig.~\ref{cooled_unpolarized_xq}. }
\label{cooled_polarized_x_Delta_q}
\end{figure}

\begin{figure}[ht]
$$
\BoxedEPSF{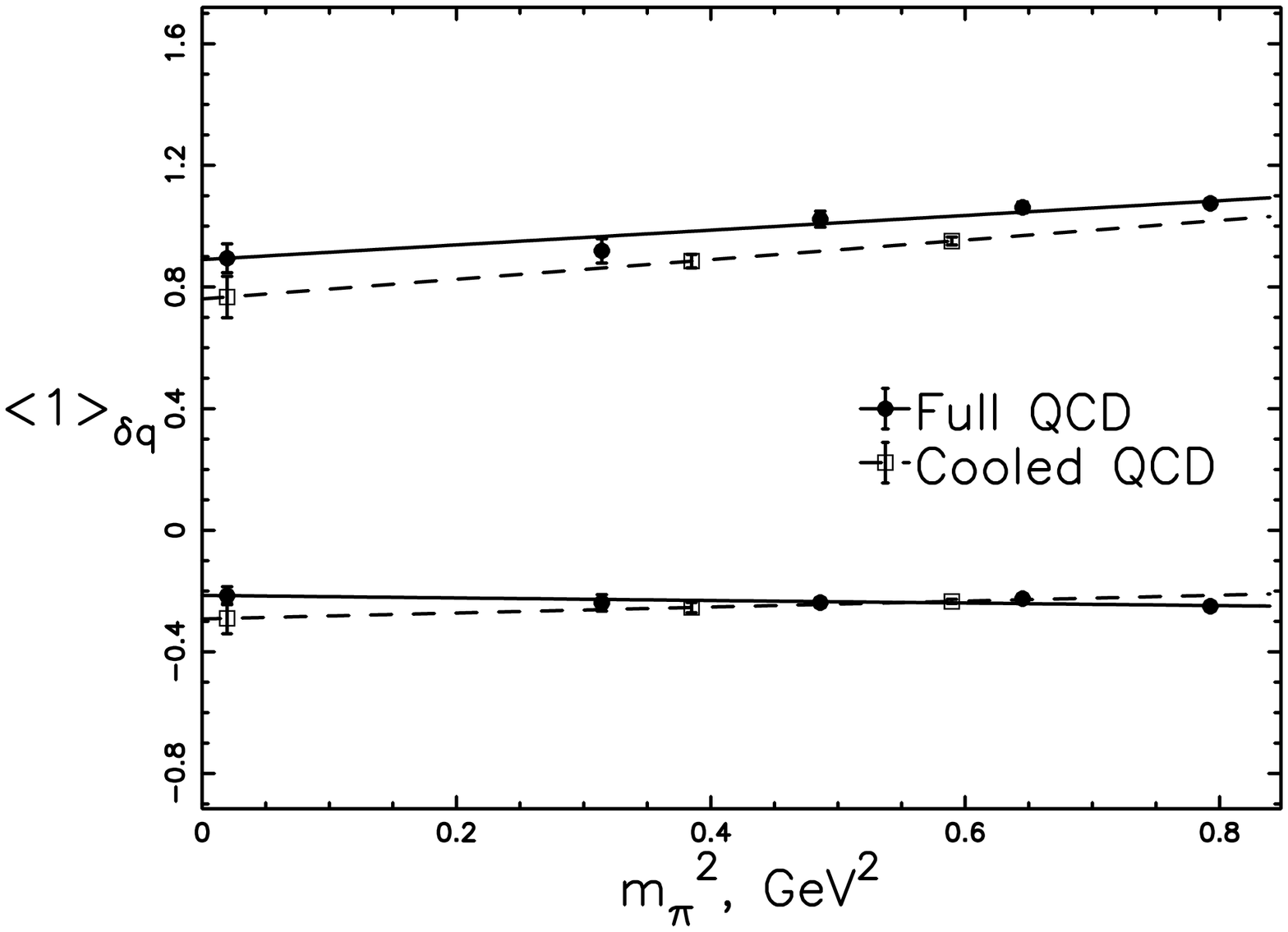 scaled 600}  
$$
\caption{Comparison of the tensor charge $\langle 1
\rangle _{\delta q}$   calculated in full QCD and using cooled configurations, as
in Fig.~\ref{cooled_unpolarized_xq}.}
\label{cooled_transversity_delta_q}
\end{figure}

\begin{figure}[ht]
$$
\BoxedEPSF{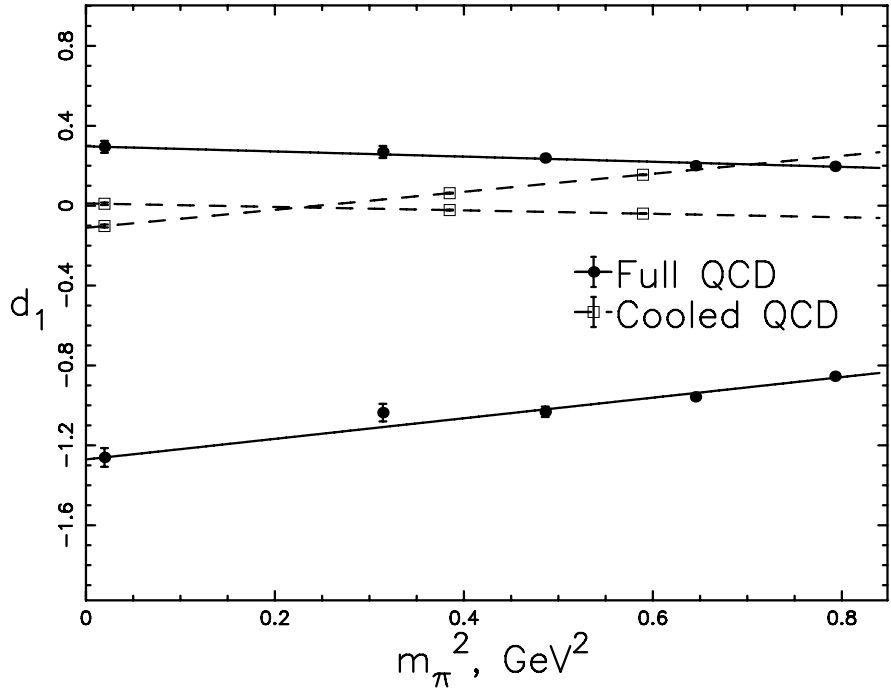 scaled 600}  
$$
\caption{Comparison of the twist-three operator $d_1$  from Eq.~\ref{eqn_dn}
calculated in full QCD and using cooled configurations, as in
Fig.~\ref{cooled_unpolarized_xq} . }
\label{cooled_d1}
\end{figure}

The moments of quark distributions calculated in cooled configurations are
tabulated in Table~\ref{cooled _data}. In addition, extrapolations in full and
cooled QCD  for the operators that can be calculated with high statistics in a
nucleon state at zero momentum, 
$\langle x \rangle _q^{(b)}$, $\langle 1 \rangle _{\Delta q}$,  
$\langle x \rangle _{\Delta q}^{(a)}$,  $\langle 1 \rangle _{\delta q}$, and
$d_1$,  are compared in Figs.~\ref{cooled_unpolarized_xq},
\ref{cooled_polarized_Delta_q}, \ref{cooled_polarized_x_Delta_q},
\ref{cooled_transversity_delta_q}, and \ref{cooled_d1} respectively. 

It is striking that the extrapolated cooled and full QCD results agree
so closely for all the twist two operators that correspond to moments of
quark distributions. This detailed agreement, generally within error
bars but always within two standard deviations, provides strong support for the
physical picture that the propagation of light quarks in the nucleon is strongly
dominated by the instanton content of the gluon configurations. It is interesting
that in the cases of $\langle x \rangle _q^{(b)}$  and $\langle x \rangle
_{\Delta q}^{(a)}$, the cooled and uncooled results agree best in the region of
light quark masses, where we expect zero mode dominance to be most
pronounced, and differ much more significantly at heavy quark masses where
they have no reason to agree in detail.

Having observed consistency between cooled and full QCD results in the cases
above where one expects agreement on the basis of instanton physics, it is
also interesting that
cooled and full QCD results differ by an order of magnitude
for the twist-three operators
$d_1$ and $d_2$ where we expect them to disagree dramatically because of
operator mixing. Recall that in connection with Eq~\ref{eqn_dn}, we pointed out that
for Wilson fermions, the operator 
$\gamma_5\gamma_{[\sigma}{\stackrel{\,\leftrightarrow}{D}}_{\{\mu_1]}
\cdots {\stackrel{\,\leftrightarrow}{D}}_{\mu_n\}}$ mixes with the lower
dimension operator
$\frac{1}{a} \gamma_5\gamma_{[\sigma}\gamma_{\{\mu_1]} \cdots
{\stackrel{\,\leftrightarrow}{D}}_{\mu_n\}}$. Hence, one expects that our lattice
measurements of $d_1$ and $d_2$ are contaminated to a large extent
by operator mixing, and indeed a calculation of the non-perturbative
mixing\cite{bib-qcdsf-d2} yields a large change in the extracted value of $d_2$. As
argued above, cooling removes the short wavelength fluctuations responsible for
renormalization and mixing, so as a result we would expect the mixing to be
reduced essentially to zero. This is precisely what is observed in
Fig.~\ref{cooled_d1} and in Table~\ref{cooled _data} where the cooled
measurements extrapolate to values close to zero and the full QCD measurements
are an order of magnitude larger.

%

\subsection{Comparison with Phenomenology}

This section presents the evidence that linear chiral extrapolation of
full lattice QCD results in the regime of quark masses accessible in this work  is
seriously inconsistent with experimental measurements of light cone quark
distributions in the nucleon. To make the argument as strong as possible, it is
useful to first compare our calculations with three other lattice calculations.
Hence, 
the quenched and full QCD calculations of
this work are compared with other related lattice calculations and with
phenomenology in Table  \ref{tab-summary}.

We have already commented on the comparison of our
quenched calculations with those of the QCDSF collaboration, and their
results
from refs 
\cite{bib-qcdsf-main,bib-qcdsf-a1,bib-qcdsf-spin,bib-qcdsf-d2,Gockeler:2000ja}
are tabulated in the first column to be compared with our quenched results in
the fourth column. Here, complementing the plots in 
Figs~\ref{quenched_unpolarized_x} -
\ref{quenched_polarized_x},  one sees detailed agreement of results calculated at
the same $\beta$ and $\kappa$, strongly supporting the accuracy and
consistency of both calculations.  

The second column shows the extrapolation to the continuum limit by the QCDSF
collaboration of several moments \cite{bib-qcdsf-tensor2}. The agreement of
these extrapolations with the first column clearly shows that in the case of
quenched QCD,  finite lattice spacing effects are small compared with the
discrepancies with experiment. Given the close agreement between full QCD and
quenched calculations in this regime of quark masses, there is no reason to
believe that finite lattice spacing effects are substantially larger in our full
QCD calculations than in the quenched case, so it would be unreasonable to
attribute discrepancies with experiment to finite lattice spacing effects. 

The third column shows the contributions to the axial charge
by the SESAM collaboration using the same gluon configurations but
completely different measurement technology\cite{Gusken:1999xy} to be
compared with our results in the fifth column.  As in the case of the quenched
comparisons, we believe this provides additional confirmation of the accuracy
and consistency of the present measurements.

The final column shows the moments that were calculated numerically from
phenomenological fits to the world supply of deep inelastic lepton scattering
data and other high energy scattering data. Data is conveniently accessible on
the web from each of the  major collaborations, and we have calculated unpolarized
moments using each of the unpolarized data sets,  CTEQ\cite{bib-cteq},
GRV\cite{bib-grv}, and MRS\cite{bib-mrs} and calculated polarized moments using
both of the sets GRSV\cite{bib-grsv}, and
GS\cite{bib-gs}.
Unfortunately, it is difficult to provide quantitative estimates of
systematic or statistical errors. Experimental data is available only over part of
the necessary range of momentum fraction, $x$,  and model assumptions are
invoked to parameterize parton distributions at large $x$ consistent with
known sum rules and physical constraints. In addition, no error correlation
matrix is provided, so it is not even possible to calculate the error in a moment
that arises from the known statistical errors in the measurements.
In order to
get some indication of the possible errors, we have calculated the moments
using each of the unpolarized or polarized data sets. In the table, we tabulate the
average value and give the maximum difference between values for all the relevant
data sets in parentheses. 
Note that these differences are small compared to the statistical errors in the
corresponding lattice measurements,   so we believe the phenomenological
uncertainties are small compared to the discrepancies with lattice extrapolations
discussed below.  Since, as argued previously, the disconnected diagrams do not
contribute to the flavor non-singlet combination
$\langle{\cal O} \rangle_{u-d} $, we have tabulated the differences between
the up and down quark contributions for the relevant combination of $q \pm
\bar q$ for comparison with the lattice calculations.

Table  \ref{tab-summary}  reveals a clear discrepancy between linearly
extrapolated lattice calculations and phenomenology.  The momentum fraction 
$\langle x \rangle _q$ is a fundamental property specifying the fraction of the
total momentum carried by a particular quark flavor. The non-singlet
momentum fraction is overestimated by more that 50\%, with full QCD yielding
0.25 - 0.29 compared with the phenomenological result 0.15. The second and
third moments are similarly overestimated by linear extrapolation.

Another important quantity is the nucleon axial charge,  $\langle 1 \rangle
_{\Delta u -\Delta d}$ governing $\beta$ decay.  Here again, one finds major
discrepancies with linear chiral extrapolation, with quenched or unquenched
calculations yielding 1.0 - 1.15 compared with the experimental value 1.26,
corresponding to a discrepancy of 10 to 25\%.  

Since these and other comparisons in Table  \ref{tab-summary} show a clear
discrepancy between phenomenology and linear chiral extrapolations and
because of the evidence summarized previously for the consistency and
accuracy of the lattice calculations at heavy quark masses, we believe that the
fault lies with the linear chiral extrapolation as discussed in the next section.

\subsection{Chiral Extrapolation}

Superficially, one could imagine that since deep inelastic scattering involves large
$Q^2$, it might be dominated by short distance behavior of the nucleon wave
function and not be strongly influenced by the long distance pion cloud. However, as
we have seen, the operator product expansion relates moments of structure
functions to a tower of local operators to be evaluated in the nucleon ground state.
Physically it is clear that the pion cloud should play a major role in
the nucleon matrix elements of these operators, and that
these contributions are strongly suppressed by the heavy quark masses and small
spatial volumes in which we have been forced to work. Whereas there is clearly
a heavy quark regime in which matrix elements vary nearly linearly with the
quark mass or $m_{\pi}^2$,  the behavior becomes highly non-linear at quark
masses sufficiently light that a substantial pion cloud is produced. 

Complementary to the linear heavy quark regime, there is a second regime
near the chiral limit in which the behavior is again simple and is specified by
chiral perturbation theory.  Here, the physics is described by an effective chiral
theory based on the would-be Goldstone bosons, and the leading non-analytic
behavior can be calculated uniquely. For example, of direct relevance to this
work, the leading non-analytic behavior of $\langle x ^n \rangle _{u-d}$ 
is\cite{Chen:2001eg,Arndt:2001ye} 
\begin{equation}
\langle x ^n \rangle _{u-d} \sim a_n \Bigl[ 1 -
{{(3 g_A^2+ 1) m_{\pi}^2} \over (4 \pi f_{\pi})^2} \ln m_{\pi}^2 \Bigr].
\end{equation}

Unfortunately, at present, there is no full analytical theory joining the chiral
regime with the heavy quark regime of the present lattice calculation. Hence
to explore  the chiral extrapolation of lattice QCD calculations to
the physical region, it is useful to use a physically motivated extrapolation
formula incorporating the correct behavior in the chiral limit. We therefore fit
the lattice data with the extrapolation formula of
Ref.~\cite{Detmold:2001jb}, in which a phenomenological cutoff $\mu$ is
introduced in the nonanalytic term to specify the size of the source
generating the pion cloud and the usual analytic term in $m_{\pi}^2$ is
included:

\begin{equation}
\label{extrapform}
\langle x ^n \rangle _{u-d} \sim a_n \Bigl[ 1 -
{{(3 g_A^2+ 1) m_{\pi}^2} \over (4 \pi f_{\pi})^2} \ln \Bigl( {m_{\pi}^2\over
m_{\pi}^2 + \mu^2} \Bigr) \Bigr] + b_n m_{\pi}^2 .
\end{equation}

Physically, it is reasonable that momenta in the pion loop should not become
infinitely large as they would in the presence of a point source, but rather be
cut off at a scale characteristic of the size of the valence quark core of the
nucleon, and alternative parameterizations in terms of a form factor give
equivalent results.

\begin{figure}[ht]
$$
\BoxedEPSF{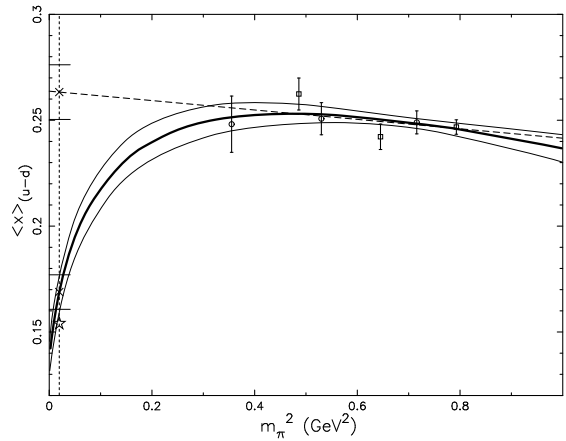 scaled 400}  
$$
\caption{
Chiral extrapolation of the momentum fraction  $\langle x  \rangle _{u-d}$ using 
Eq.~\ref{extrapform}. Full QCD and quenched data calculated in this present work are
denoted by squares and circles respectively, and the phenomenological result is
indicated by the star. The least-squares fit and jackknife error bars are denoted by
the heavy solid line and surrounding light lines. }
\label{chiral_extrapolation_a}
\end{figure}

The result of using this extrapolation formula for 
the difference between the up and down quark momentum fraction $\langle x
^n \rangle _{u-d}$  is shown in 
Fig.~\ref{chiral_extrapolation_a}. Although this extrapolation is only valid in full QCD,
since full QCD and quenched results are equivalent in the regime of our
calculations,  to improve statistics, we have also included our quenched results.
The heavy solid curve is the result of a least-squares fit of $a_1$ and $b_1$  in
Eq.~\ref{extrapform} with fixed $\mu=550 $MeV and the 
the light solid lines indicate the  jackknife error band. Here we see clearly that  the
extrapolation formula containing the leading chiral behavior
is consistent both with the lattice measurements and the experimental data. The
fact that the effect of the chiral logarithm is much larger in this matrix element
than in more familiar mass measurements is again a manifestation of the
variational theorem. Making an order $\epsilon$ error in the pion cloud of the
wave function makes an order $\epsilon$ error in matrix elements of general
operators but only an order $\epsilon^2$ error in the mass. The obvious problem
with the present argument is that although a cutoff of order $0.3$ fm is physically
reasonable, it has not been calculated from first principles and may be regarded as
a single free parameter introduced to fit a single experimental measurement.

\begin{figure}[ht]
$$
\BoxedEPSF{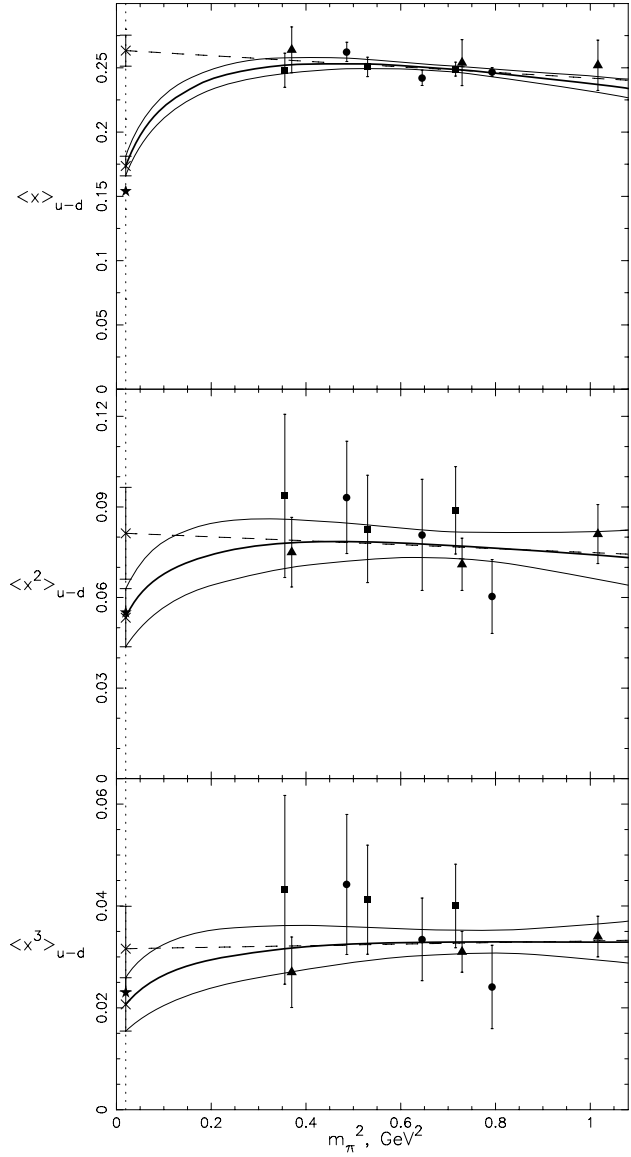 scaled 600}  
$$
\caption{
Chiral extrapolation of the first three moments of
the proton quark distribution,  $\langle x ^n \rangle _{u-d}$, using 
Eq.~\ref{extrapform}.
Full QCD and quenched data calculated in this present work are
denoted by diamonds and squares  respectively, QCDSF quenched 
data\protect\cite{bib-qcdsf-main} are
denoted by triangles, and the phenomenological results are indicated by stars.
The least-squares fit and jackknife error bars are denoted by
the heavy solid line and surrounding light lines.}

\label{chiral_extrapolation_b}
\end{figure}

To see that this single cutoff parameter resolves the discrepancy with experiment
for a number of observables,  the results of using Eq.~\ref{extrapform} for the first
three moments of the difference between the up and down quark density
distributions
$\langle x ^n \rangle _{u-d}$ are shown in figure \ref{chiral_extrapolation_b}.
Because of the larger error bars in the higher moments, we have also included 
the quenched QCDSF results\cite{bib-qcdsf-main} to
provide a larger lever arm in the least squares fit. As seen in the figure, the
single value of $\mu$ provides simultaneous agreement with experiment for
all three moments. As shown in Ref.~\cite{Detmold:2001jb}, this strong
chiral behavior of the three lowest moments is also observed in chiral bag
models.  Furthermore, similar extrapolation with a comparable cutoff also
reconciles the strong discrepancy between linear extrapolation of lattice
results and experimental nucleon magnetic moments\cite{Leinweber:1998ej}.
Thus, a consistent picture is emerging concerning the importance of large
effects of the pion cloud in chiral extrapolations. Unfortunately, because
spin-dependent structure functions also involve significant contributions
from Delta excitations and chiral perturbation theory becomes less well
controlled, we do not presently have a corresponding physical interpolation
formula for
$\langle x ^n\rangle _{\Delta u - \Delta d}$.

The curves in Fig.~\ref{chiral_extrapolation_b} suggest that precision
measurements down to $ m_{\pi}^2 = 0.05$~GeV$^2$ are required to determine
the parameters of the chiral extrapolation and thereby provide reliable
extrapolation of moments of quark distributions.
 The computational resources required for such calculations may be estimated
using 
the cost function~\cite{Lippert:zq} obtained by the SESAM collaboration, which
provided the gluon configurations.  For present purposes, the number of floating
point operations per independent gluon configuration,
$N$ in Teraflops-years, may be conveniently
written
$$   N \simeq  .038 \biggl[\frac{L}{4}\biggr]^{4.55}
\biggl[\frac{.08}{a}\biggr]^{7.25}
\biggl[\frac{.3}{\frac{m_{\pi}}{m_{\rho}}}\biggr]^{2.7}.
$$
Because the spatial derivatives and non-zero momentum projections
required to calculate moments of structure functions require high
Monte Carlo statistics, it is necessary to calculate of the order of 400
independent configurations. Including equilibration and calculation
at higher quark masses, the total computer time is approximately
twice that required for 400 configurations at the lowest quark
mass. Hence, a calculation with a  lattice spacing $a =
0.1\;$fm and $\frac{m_{\pi}}{m_{\rho}}= 0.3$ would require approximately 8
Teraflops-years,  i.e.,  dedicated use of a computer that
sustains 8 Teraflops on QCD for one year. Such resources should become
available in the next generation of lattice QCD computers.

\section{Summary and Conclusions}

This work has presented the first calculation of the moments of light cone quark
distributions in full QCD. The methodology has been presented in detail and
validated by numerous consistency checks and comparison with other relevant
lattice calculations.

One major result of this work is the close agreement of full QCD and quenched
calculations for quark masses corresponding to $m_{\pi}$ above 500 MeV. This
agreement rules out the conjecture that discrepancies with experiment could be
attributed to quenching effects.

A second salient result is clear evidence that linear extrapolation of full QCD
lattice results from pion masses above 500 MeV is in serious disagreement with
experiment, ranging from the order of 50\% for the quark momentum fraction
to 10-25\% for the axial charge.

Motivated by the inadequacy of linear chiral extrapolation, we have shown that
extrapolation including the leading nonanalytic behavior of chiral perturbation
theory has the potential to yield results consistent with phenomenology. We
have explicitly shown that an extrapolation formula with a single
phenomenological cutoff simultaneously fits the first three moments of the
quark momentum fraction. 

Finally, we have shown qualitative agreement between full QCD and cooled
lattice configurations retaining only the contributions of instantons, providing
additional evidence for the role of instantons in light hadron structure and of
zero-mode dominance.

This work points the way for a number of promising steps in our continued
quest to understand hadron structure from first principles using lattice QCD. 
One should clearly undertake a systematic program of using a partially
quenched chiral expansion for extrapolation and measure the relevant
parameters of the effective chiral theory. To complement the flavor
non-singlet matrix elements of this work, we need to calculate the disconnected
diagrams required to compare flavor singlet matrix elements with
phenomenology.  The continuum limit of the SESAM results should be explored
by supplementing the existing SCRI configurations with additional quark masses
so that calculations at $\beta = 5.6, 5.5$, and 5.3 may be compared and
extrapolated.
It is desirable to undertake nonperturbative
renormalization without Gribov ambiguities associated with gauge fixing,
and
to this end we note that the Schr\"odinger functional method has now been
used to calculate the average quark momentum in the pion
\cite{Guagnelli:2000em,Jansen:2000xm,Guagnelli:1999gu,Guagnelli:1998ve}.
Quenched calculations with chiral fermions in a spatial volume of
3.2 fm should be carried out to extend the pion mass down to 250 MeV and to
remove the problem of operator mixing in the twist-3 matrix elements.  Finite
volume formulae for operator matrix elements should be derived to correct
residual finite volume effects. As lattice results come closer to phenomenology,
it will  be important to have quantitative understanding of the errors in
moments of structure functions arising from systematic errors, as well as the
error correlation matrix describing statistical errors. Finally, with the advent of
multi-Terascale computers, the promise of full QCD calculations with light pion
masses will finally be within our grasp.

We note that subsequent to the completion of this work, a thorough analysis of
the error correlations in the experimental measurements of polarized parton
distributions was performed in ref~\cite{Blumlein:2002be}, with the result that
the errors in the phenomenological values of moments of $\Delta u -\Delta d$ in 
table~\ref{tab-summary} are of the order of ten percent.

\acknowledgments

The authors wish to
thank R.Horsley, W. Melnitchouk, Edward Shuryak, and A. W. Thomas for helpful
discussions and communications.
The research in this work was supported in part by the U.S.
Department of Energy under cooperative research agreement
DE-FC02-94ER40818 and contracts
DE-FG02-91ER40676 and DE-FG02-97ER41022. 
R.~G.~E. was supported by DOE contract DE-AC05-84ER40150 under which the
Southeastern Universities Research Association (SURA) operates the
Thomas Jefferson National Accelerator Facility (TJNAF).
S.~C., Th.~L., and K.~S.
acknowledge the EU Network HPRN-CT-2000-00145 ``Hadron
Phenomenology from Lattice QCD" for support.
We gratefully acknowledge use of computer resources
provided by the MIT and Jefferson Lab LHPC clusters, the SCRI CM-2, and the Jlab
QCDSP.  We thank the John von Neumann-Institute for Computing for
providing  the Cray T3E and APE100 facilities and the staff of the
computer centers ZAM  at FZ-JŸlich and DESY/Zeuthen for their enormous
support.

\appendix

\section{Calculation of Sequential Propagators}

This appendix describes the explicit calculation of the sequential nucleon sources
used in the calculations in this work.

Denote the smeared field by $\Psi$, the original unsmeared field by
$\psi$ and the gauge invariant smearing function defined by
Eq.~\ref{opt-psi1Hpsi} by $F$ so that
\begin{equation}
\Psi^a_\alpha(\vec{x}_0,t_0) = \int d^3x \,
F^{a a^\prime}_{\alpha\alpha^\prime}(\vec{x}_0,\vec{x}) \,
\psi^{a^\prime}_{\alpha^\prime}(\vec{x}, t_0).
\end{equation}
The forward propagator from a smeared source to a point sink, denoted
$$
S^{a a^\prime}_{\alpha\alpha^\prime}(\psi\Psi|\vec{x},t;\vec{x}_0, t_0) 
\stackrel{\text{def}}{=} 
\langle \psi^a_\alpha(\vec{x},t)\,\bar{\Psi}^{a^\prime}_{\alpha^\prime}
(\vec{x}_0,t_0)\rangle,
$$
is the solution to the linear system
\begin{equation}
\label{DSF}
\int dt^\prime\,d^3x^\prime\, 
{D_W}^{a a^{\prime\prime}}_{\alpha{\alpha^{\prime\prime}}}
(\vec{x},t;\vec{x}^\prime,t^\prime)
S^{a^{\prime\prime} a^\prime}_{{\alpha^{\prime\prime}}\alpha^\prime}
(\psi\Psi|\vec{x}^\prime,
t^\prime;\vec{x}_0, t_0)=\delta(t-t_0)\,F^{a a^\prime}_{\alpha\alpha^\prime}
(\vec{x},\vec{x}_0),
\end{equation}
where  $D_W$ denotes the Wilson Dirac operator and the smearing function is
the source term. The propagator from a smeared source to a smeared sink is
obtained by an additional smearing,
\begin{eqnarray}
\label{SFS}
S^{a a^\prime}_{\alpha\alpha^\prime}(\Psi\Psi|\vec{x},t;\vec{x}_0, t_0) 
&\stackrel{\text{def}}{=}& 
\langle \Psi^a_\alpha(\vec{x},t)\,
\bar{\Psi}^{a^\prime}_{\alpha^\prime}(\vec{x}_0,t_0)\rangle\nonumber\\
&=& \int\,d^3x^\prime\, F^{a a^{\prime\prime}}_{\alpha{\alpha^{\prime\prime}}}
(\vec{x},\vec{x}^\prime)\,
S^{a^{\prime\prime} a^\prime}_{{\alpha^{\prime\prime}}\alpha^\prime}
(\psi\Psi|\vec{x}^\prime,t;\vec{x}_0, t_0).
\end{eqnarray}

The  three point function specifying $u$ quark operator matrix elements,
using the  proton current, Eq.~\ref{app-nucl1},  with smeared quark fields, 
may be written:
\begin{eqnarray}
\langle {J}^\alpha &O& \bar{{J}}^{\alpha^\prime} \rangle^{(u)}(t_i,t_f,t_o, 
\vec{x}_i)=
\int d^3x_o\,\int d^3x_f e^{i\vec{p}\vec{x}_f} 
\epsilon^{abc}\epsilon^{a^\prime b^\prime c^\prime} \Gamma_{\beta\gamma} 
\bar{\Gamma}_{\beta^\prime\gamma^\prime} \times \\
&&\langle U^a_\alpha(\vec{x}_f, t_f) U^b_\beta(\vec{x}_f, t_f)  
D^c_\gamma(\vec{x}_f, t_f)\,
\bar{u}^d_\nu(\vec{x_o},t_o) O_{\nu\nu^\prime}^{d
d^\prime}u^{d^\prime}_{\nu^\prime} (\vec{x}^\prime_o, t^\prime_o)\,
\bar{U}^{a^\prime}_{\alpha^\prime}(\vec{x}_i, t_i) 
\bar{U}^{b^\prime}_{\beta^\prime}(\vec{x}_i, t_i)  
\bar{D}^{c^\prime}_{\gamma^\prime}(\vec{x}_i, t_i)\rangle \nonumber
\end{eqnarray}
where $t_i$, $t_f$ and $t_o$ are time coordinates of the source, sink and 
operator insertion, $u$ and $U$ denote point and smeared fields  for the
$u$-quark,  $d$ and $D$ denote point and smeared fields for the $d$-quark, and
$\Gamma=C\gamma_5$. The three point 
function for a $d$ quark operator has the same form with the operator insertion
$\bar{d}^d_\nu(\vec{x_o},t_o) O_{\nu\nu^\prime}^{d
d^\prime}d^{d^\prime}_{\nu^\prime} (\vec{x}^\prime_o, t^\prime_o)$.

After performing the connected diagram contractions, the $u$ and $d$ operator
matrix elements may be written:
\begin{eqnarray}
\label{MSOM}
\langle {J}^\alpha &O& \bar{{J}}^{\alpha^\prime} \rangle^{(u,d)}(t_i,t_f,t_o) 
=  \int d^3x_o\,\int d^3x_f e^{i\vec{p}\vec{x}_f}
{\cal M}^{(u,d)\,[\alpha\alpha^\prime]a a^\prime}_{\mu\mu^\prime}
(\vec{x}_f, t_f, \vec{x}_i , t_i)\,\\
&\times&S^{ad}_{\mu\nu}(\Psi\psi|\vec{x}_f,t_f;\vec{x}_o, t_o)\,
O_{\nu\nu^\prime}^{d d^\prime} S^{d^\prime a^\prime}_{\nu^\prime\mu^\prime}
(\psi\Psi|\vec{x}^\prime_o, t^\prime_o, \vec{x}_i,t_i)\nonumber
\end{eqnarray}
where 
\begin{eqnarray}
\label{MSS}
&{\cal M}&^{(u)\,[\alpha\alpha^\prime]a a^\prime}_{\mu\mu^\prime}(\vec{x}_f, 
t_f, t_i) \stackrel{\text{def}}{=}  
\epsilon^{abc}\epsilon^{a^\prime b^\prime c^\prime}\,
S^{c c^\prime}_{\gamma\gamma^\prime}\times\\
&&\times\biggl[ S^{b b^\prime}_{\alpha\alpha^\prime}
\Gamma_{\mu\gamma}\bar{\Gamma}_{\mu^\prime\gamma^\prime} +
S^{b b^\prime}_{\alpha\beta^\prime}\Gamma_{\mu\gamma}
\bar{\Gamma}_{\beta^\prime\gamma^\prime}\delta_{\mu^\prime\alpha^\prime} +
S^{b b^\prime}_{\beta\alpha^\prime}\Gamma_{\beta\gamma}
\bar{\Gamma}_{\mu^\prime\gamma^\prime}\delta_{\mu\alpha} +
S^{b b^\prime}_{\beta\beta^\prime}\Gamma_{\beta\gamma}
\bar{\Gamma}_{\beta^\prime\gamma^\prime}\delta_{\mu\alpha}
\delta_{\mu^\prime\alpha^\prime}\biggr],
\nonumber\\
\nonumber\\
&{\cal M}&^{(d)\,[\alpha\alpha^\prime]a a^\prime}_{\mu\mu^\prime}(\vec{x}_f, 
t_f, t_i) \stackrel{\text{def}}{=} 
\epsilon^{abc}\epsilon^{a^\prime b^\prime c^\prime}\, 
\biggl[ S^{b b^\prime}_{\alpha \alpha^\prime} 
S^{c c^\prime}_{\beta \beta^\prime} +
S^{b b^\prime}_{\alpha \beta^\prime} 
S^{c c^\prime}_{\beta \alpha^\prime}
\biggr] \Gamma_{\beta \mu} \Gamma_{\beta^\prime \mu^\prime},
\end{eqnarray}
and $S^{a a^\prime}_{\mu\mu^\prime} \equiv
S^{a a^\prime}_{\mu\mu^\prime}(\Psi\Psi|\vec{x}_f,t_f,\vec{x}_i,t_i) $. 

We now define the backward propagator 
\begin{equation}
\label{BDEF}
{\cal B}^{[\alpha\alpha^\prime]d a^\prime}_{\nu^{\prime\prime}\mu^\prime}
(\vec{x}_o, t_o,t_f,\vec{x}_i,t_i)\stackrel{\text{def}}{=} \int d^3x_f\, e^{-i\vec{p}
\vec{x}_f}\, S^{da}_{\nu^{\prime\prime}\mu^{\prime\prime}}(\psi\Psi|
\vec{x}_o, t_o;\vec{x}_f,\vec{x}_i,t_f)
\,\gamma_{\mu^{\prime\prime}\mu}^5\,{\cal M}^{\ast[\alpha\alpha^\prime]a 
a^\prime}_{\mu\mu^\prime}(\vec{x}_f, t_f, t_i).
\end{equation}

Since it propagates from the sink to the operator, we
use the relation  $\gamma^5 S(x,y) \gamma^5 = S^\dagger(y,x)$
to obtain
\begin{equation}
\biggl[\int d^3x_f\, e^{i\vec{p}\vec{x}_f}\,{\cal M}^{[\alpha\alpha^\prime]a 
a^\prime}_{\mu\mu^\prime}(\vec{x}_f, t_f,\vec{x}_i, t_i)\,
S^{ad}_{\mu\nu}(\Psi\psi|\vec{x}_f,t_f;\vec{x}_o, t_o)\biggl]^\ast=
\gamma_{\nu\nu^{\prime\prime}}^5 {\cal B}^{[\alpha\alpha^\prime]
da^\prime}_{\nu^{\prime\prime}\mu^\prime}(\vec{x}_o, t_o,t_f, \vec{x}_i,t_i)
\end{equation}
so that (\ref{MSOM}) becomes
\begin{equation}
\label{BOS}
\langle {J}^\alpha O {J}^{\alpha^\prime} \rangle(t_i,t_f,t_o) = \int d^3x_o\,
\gamma_{\nu\nu^{\prime\prime}}^5 {\cal B}^{\ast[\alpha\alpha^\prime]
da^\prime}_{\nu^{\prime\prime}\mu^\prime}(\vec{x}_o, t_o,t_f, \vec{x}_i,t_i)\,
O_{\nu\nu^\prime}^{dd^\prime} S^{d^\prime a^\prime}_{\nu^\prime\mu^\prime}
(\psi\Psi|\vec{x}_o, t_o, \vec{x}_i, t_i).
\end{equation}

As in Eq.~(\ref{DSF}), the backward 
propagator ${\cal B}^{[\alpha\alpha^\prime]}$ defined by (\ref{BDEF}) 
may be calculated by solving the linear system:
\begin{eqnarray}
\label{DSFMSM}
\int dt^\prime\,d^3x^\prime\, &{D_W}&^{a d}_{\mu^{\prime\prime}
\nu^{\prime\prime}}(\vec{x},t;\vec{x}_o,t_o)
{\cal B}^{[\alpha\alpha^\prime]da^\prime}_{\nu^{\prime\prime}\mu^\prime}
(\vec{x}_o,t_o,t_f)\\
&=&\delta(t-t_f)\,
e^{-i\vec{p}\vec{x}}\,\int d^3x^\prime\, 
F^{a a^{\prime\prime}}_{\mu^{\prime\prime}\rho}(\vec{x},\vec{x}^\prime) 
\gamma_{\rho\mu}^5 {\cal M}^{\ast[\alpha\alpha^\prime]a^\prime\prime 
a^\prime}_{\mu\mu^\prime}(\vec{x}^\prime, t_f).\nonumber
\end{eqnarray}

Finally, in the same notation, the momentum 
projected two-point function with smeared sources is :
\begin{eqnarray}
\label{JJUU}
\langle {J}^\alpha \bar{{J}}^{\alpha^\prime} \rangle(t_i,t_f,\vec{x}_i)
&=& \int d^3x_f e^{i\vec{p}\vec{x}_f} 
\epsilon^{abc}\epsilon^{a^\prime b^\prime c^\prime} \Gamma_{\beta\gamma} 
\bar{\Gamma}_{\beta^\prime\gamma^\prime} \times \\
&&\langle U^a_\alpha(\vec{x}_f, t_f) U^b_\beta(\vec{x}_f, t_f)  
D^c_\gamma(\vec{x}_f, t_f)\,
\bar{U}^{a^\prime}_{\alpha^\prime}(\vec{x}_i, t_i) 
\bar{U}^{b^\prime}_{\beta^\prime}(\vec{x}_i, t_i)  
\bar{D}^{c^\prime}_{\gamma^\prime}(\vec{x}_i, t_i)\rangle
\nonumber\\
&=&\int d^3x_f e^{i\vec{p}\vec{x}_f}
\epsilon^{abc}\epsilon^{a^\prime b^\prime c^\prime} \Gamma_{\beta\gamma} 
\bar{\Gamma}_{\beta^\prime\gamma^\prime} 
S^{c c^\prime}_{\gamma\gamma^\prime} \biggl[S^{a
a^\prime}_{\alpha\alpha^\prime}  S^{b b^\prime}_{\beta\beta^\prime} - 
S^{a b^\prime}_{\alpha\beta^\prime} 
S^{b a^\prime}_{\beta\alpha^\prime}\biggr].\nonumber
\end{eqnarray}

\newpage

\begin{table}

\begin{tabular}{llllllll}
Data set        & QCD & $L_x^3\times L_t$ & $\beta$ & $\kappa_{sea}$ 
& $\kappa_{val}$ & Approx. & Trajectory \\ 
 & & & & &  & \# configs. & separation \\
\tableline
SESAM        & full & $16^3\times 32$ & 5.6 & 0.1560 & 0.1560
& 200  & 25 \\
             &      &                &     & 0.1565 &   0.1565              
& 200   & 25 \\
             &      &                &     & 0.1570 &         0.1570        
& 200  & 25  \\
             &      &                &     & 0.1575 &       0.1575          
& 200  & 25  \\
SCRI         & full & $16^3\times 32$ & 5.5 & 0.1596 & 0.1596 
& 100   & 20 \\
             &      &                &     & 0.1600 &     0.1600             
& 100   & 20 \\
             &      &                &     & 0.1604 &     0.1604            
& 100   & 18 \\
MIT       & quenched &  $16^3\times 32$ & 6.0 & $m_q =\infty$ & 0.1530     
& 200 &\\
             &          &                 &     &        & 0.1540     
& 200 & \\
             &          &                 &     &        & 0.1550     
& 200 & \\
SESAM-cooled   & cooled & $16^3\times 32$ & 5.6 & 0.1560 & 0.1235 & $100$
& \\
             &     &                 &     & 0.1570 & 0.1246     
& 100 & \\

\end{tabular}

\caption{Parameters specifying the full QCD, quenched, and cooled gluon
configurations used in calculations of moments of quark distributions.
For full QCD calculations, the number of  hybird Monte Carlo trajectories
between measurements is given in the last column.}
\label{t_datasets}
\end{table}

\begin{table}[t]

\begin{tabular}{lccrl}
observable & H(4) & mixing & $\vec{P}$ & lattice operator\\
\tableline
$\langle x \rangle _q^{(a)}$ & {\bf 6}$_3^+$ & no & $1$  & $\bar{q}
\gamma_{\{1}  {\stackrel{\,\leftrightarrow}{D}}_{4\}} {q}$  \\
$\langle x \rangle _q^{(b)}$ & {\bf 3}$_1^+$ & no & 0& $\bar{q} \gamma_{4} 
{\stackrel{\,\leftrightarrow}{D}}_{4} {q} - \frac{1}{3}(\bar{q} \gamma_{1} 
{\stackrel{\,\leftrightarrow}{D}}_{1} {q}
+\bar{q} \gamma_{2} {\stackrel{\,\leftrightarrow}{D}}_{2} {q}+\bar{q} 
\gamma_{3} {\stackrel{\,\leftrightarrow}{D}}_{3} {q})$  \\
$\langle x ^2 \rangle _q$     & {\bf 8}$_1^-$ & yes & $1$ & $\bar{q}
\gamma_{\{1}  {\stackrel{\,\leftrightarrow}{D}}_{1} 
{\stackrel{\,\leftrightarrow}{D}}_{4\}} {q} - \frac{1}{2} \bar{q} (\gamma_{\{2} 
{\stackrel{\,\leftrightarrow}{D}}_{2} 
{\stackrel{\,\leftrightarrow}{D}}_{4\}}+ \gamma_{\{3} 
{\stackrel{\,\leftrightarrow}{D}}_{3} 
{\stackrel{\,\leftrightarrow}{D}}_{4\}}){q}$ \\
$\langle x^3 \rangle _q$     & {\bf 2}$_1^+$ & no$^*$ & $1$ & $\bar{q}
\gamma_{\{1}  {\stackrel{\,\leftrightarrow}{D}}_{1} 
{\stackrel{\,\leftrightarrow}{D}}_{4} 
{\stackrel{\,\leftrightarrow}{D}}_{4\}} {q}
+ \bar{q} \gamma_{\{2} {\stackrel{\,\leftrightarrow}{D}}_{2} 
{\stackrel{\,\leftrightarrow}{D}}_{3} 
{\stackrel{\,\leftrightarrow}{D}}_{3\}} {q} 
- (\,3\,\leftrightarrow\,4\,)$ \\
$\langle 1 \rangle _{\Delta q}$     & {\bf 4}$_4^+$ & no & 0  & $\bar{q} 
\gamma^{5} \gamma_{3} {q}$  \\
$\langle x \rangle _{\Delta q}^{(a)}$ & {\bf 6}$_3^-$  & no & $1$ & $\bar{q} 
\gamma^5 \gamma_{\{1} {\stackrel{\,\leftrightarrow}{D}}_{3\}} {q}$ \\
$\langle x \rangle _{\Delta q}^{(b)}$ & {\bf 6}$_3^-$  & no & 0 & $\bar{q}
\gamma^5 
\gamma_{\{3} {\stackrel{\,\leftrightarrow}{D}}_{4\}} {q}$ \\
$\langle x^2 \rangle _{\Delta q}$     & {\bf 4}$_2^+$  & no & $1$ & $\bar{q} 
\gamma^5 \gamma_{\{1} {\stackrel{\,\leftrightarrow}{D}}_{3} 
{\stackrel{\,\leftrightarrow}{D}}_{4\}} {q}$ \\
$\langle 1 \rangle _{\delta q}$     & {\bf 6}$_1^+$  & no & 0  & $\bar{q}
\gamma^{5} \sigma_{34}  {q}$  \\
$\langle x \rangle _{\delta q}$    & {\bf 8}$_1^-$ & no & $1$   & $\bar{q}
\gamma^{5} 
\sigma_{3\{4} {\stackrel{\,\leftrightarrow}{D}}_{1\}}{q}$  \\
$d_1$     & {\bf 6}$_1^+$  & no$^{**}$ & 0  & $\bar{q} \gamma^5 \gamma_{[3} 
{\stackrel{\,\leftrightarrow}{D}}_{4]}{q}$ \\
$d_2$     & {\bf 8}$_1^-$  & no$^{**}$ & $1$  & $\bar{q} \gamma^5
\gamma_{[1}  {\stackrel{\,\leftrightarrow}{D}}_{\{3]} 
{\stackrel{\,\leftrightarrow}{D}}_{4\}} {q}$ \\
\end{tabular}

\caption{Operators used to measure moments of quark distributions.  Different
lattice operators corresponding to the same continuum operator are denoted by
superscripts $a$ and $b$.  Subscripts of irreducible representations of H(4)
distinguish different representations of the same dimensionality and
superscripts denote charge conjugation C.  In the operator mixing column, no$^*$
indicates a case in which mixing generically could exist but vanishes
perturbatively for Wilson or overlap fermions and  no$^{**}$ indicates
perturbative mixing with lower dimension operators for Wilson fermions but no
mixing for overlap fermions. The entry in column $\vec{P}$ denotes the
number of spatial components of the nucleon momentum, $\vec{P}$, that
must be chosen non-zero. Operators requiring one non-zero component have
been written for $\vec{P}$ in the 1- direction and $\vec{S}$ in the
3-direction.}
\label{tab-operators}
\end{table}

\begin{table}[t]

\begin{tabular}{lrrrrr}
observable & $\gamma$ & $B^{LATT}$ & $B^{{\overline{MS}}}$ & $Z(\beta=6.0)$ 
& $Z(\beta=5.6)$\\
\tableline
$\langle x \rangle _q^{(a)}$  &    8/3 &  -3.16486 &     -40/9 & 0.9892 & 0.9884
\\
$\langle x \rangle _q^{(b)}$ &    8/3 &  -1.88259 &     -40/9 & 0.9784 & 0.9768
\\
$\langle x ^2 \rangle _q$ &   25/6 & -19.57184 &     -67/9 & 1.1024 & 1.1097 \\
$\langle x^3 \rangle _q$  & 157/30 & -35.35192 & -2216/225 & 1.2153 &
1.2307 \\
$\langle 1 \rangle _{\Delta q}$&      0 &  15.79628 &         0 & 0.8666 & 0.8571
\\
$\langle x \rangle _{\Delta q}^{(a)}$&    8/3 &  -4.09933 &     -40/9 & 0.9971 &
0.9969 \\
$\langle x \rangle _{\Delta q}^{(b)}$&    8/3 &  -4.09933 &     -40/9 & 0.9971 &
0.9969 \\
$\langle x^2 \rangle _{\Delta q}$&   25/6 & -19.56159 &     -67/9 & 1.1023 &
1.1096 \\
$\langle 1 \rangle _{\delta q}$ &      1 &  16.01808 &        -1 & 0.8563 & 0.8461
\\
$\langle x \rangle _{\delta q}$ &      3 &  -4.47754 &        -5 & 0.9956 & 0.9953
\\
$d_1$              &      0 &   0.36500 &         0 & 0.9969 & 0.9967 \\
$d_2$              &    7/6 & -15.67745 &    -35/18 & 1.1159 & 1.1242
\end{tabular}

\caption{Renormalization constants}
\label{tab-renorm1}
\end{table}

\begin{table}
\begin{tabular}{llllll}
 & $\beta=6/g_0^2$ & $\kappa_{crit}$ & $a_N$ (fm) 
 &  $a_\rho$ (fm) & Ref.  \\
\tableline
LANL & 5.6 & 0.1585  & 0.086(14) & 0.076(9)   
& \cite{bib-lanl-spectro}\\
     & 5.5 & 0.16145 & 0.116(7)  & 0.104(5)   
& \cite{bib-lanl-spectro}\\
     & 5.4 & 0.16450 & 0.164(13) & 0.136(9) 
& \cite{bib-lanl-spectro}\\
HEMCGC & 5.3 & 0.16794 & 0.124(4)  & 0.108(2) 
& \cite{bib-hemcgc-spectro}\\
SCRI  & 5.5 & 0.16116  & 0.109(4)  & 0.090(3) 
& \cite{bib-edwards-spectro}\\
SESAM  & 5.6 & 0.1585  & 0.091(16) & 0.076(10) 
& \cite{bib-sesam-spectro}
\end{tabular}

\caption{Parameters and  lattice spacing for published dynamical Wilson
fermion calculations. This table summarizes 
calculations on $16^3
\times 32$ lattices and the resulting determinations of the lattice spacing
from the masses of the nucleon, $a_N$, and from the $\rho$ meson,
$a_{\rho}$.}
\label{t_world-dyn-wil}
\end{table}

\begin{table}[!ht]
\centering
\begin{tabular}{|l|c|cccc|}
 & $ \kappa_c$=0.1571 & 0.1550 & 0.1540 & 0.1530 & \\
\hline
\hline
$\langle x \rangle _u^{(b)}$ & 0.454(29) & 0.458(17) & 0.464(10) & 0.465(7) & \\
$\langle x \rangle _d^{(b)}$ & 0.203(14) & 0.209(8) & 0.213(5) & 0.216(3) & \\
\hline
$\langle x \rangle _u^{(a)}$ & 0.308(125) & 0.375(64) & 0.402(47) & 0.437(41) & \\
$\langle x \rangle _d^{(a)}$ & 0.159(57) & 0.179(29) & 0.178(21) & 0.192(19) & \\
\hline
$\langle x ^2 \rangle _u$ & 0.119(61) & 0.135(32) & 0.129(23) & 0.142(20) & \\
$\langle x ^2 \rangle _d$ & 0.0286(324) & 0.0410(167) & 0.0464(118) & 0.0527(108) &
\\
\hline
$\langle x^3 \rangle _u$  & 0.0369(357) & 0.0441(192) & 0.0504(126) & 0.0532(107)
& \\
$\langle x^3 \rangle _d$  & -0.00893(1842) & 0.000957(10126) & 0.00917(638) &
0.0132(53) & \\
\hline
$\langle 1 \rangle _{\Delta u}$ & 0.888(80) & 0.915(46) & 0.926(28) & 0.939(20) & \\
$\langle 1 \rangle _{\Delta d}$  & -0.241(58) & -0.244(38) & -0.250(18) & -0.251(12) &
\\
\hline
$\langle x \rangle _{\Delta u}^{(b)}$ & 0.215(25) & 0.228(14) & 0.238(9) & 0.243(7) & \\
$\langle x \rangle _{\Delta d}^{(b)}$ & -0.0535(164) & -0.0564(97) & -0.0561(56) 
& -0.0578(39) & \\
\hline
$\langle x \rangle _{\Delta u}^{(a)}$ & 0.141(123) & 0.199(65) & 0.220(45) & 0.251(37)
& \\
$\langle x \rangle _{\Delta d}^{(a)}$ & -0.00144(7536) & -0.0200(407) & -0.0338(283)
&  -0.0415(215) & \\
\hline
$\langle x^2 \rangle _{\Delta u}$ & 0.0269(428) & 0.0514(236) & 0.0639(152) 
& 0.0758(121) & \\
$\langle x^2 \rangle _{\Delta d}$ & 0.00274(2530) & -0.00555(1431) & -0.0107(90)
 & -0.0145(66) & \\
\hline
$\langle 1 \rangle _{\delta u}$ & 1.014(82) & 1.035(49) & 1.045(28) & 1.055(20) & \\
$\langle 1 \rangle _{\delta d}$ & -0.199(46) & -0.222(27) & -0.241(16) & -0.251(11) &
\\
\hline
$\langle x \rangle _{\delta u}$ & 0.337(209) & 0.422(111) & 0.474(79) & 0.512(63) & \\
$\langle x \rangle _{\delta d}$ & -0.0694(689) & -0.0723(371) & -0.0638(253) &
-0.0682(201) & \\
\hline
$ d_1^u $ & -1.354(75) & -1.079(50) & -0.914(24) & -0.779(14) & \\
$ d_1^d $ & 0.282(42) & 0.230(28) & 0.208(13) & 0.183(8) & \\
\hline
$ d_2^u $ & -0.233(86) & -0.189(52) & -0.172(29) & -0.150(19) & \\
$ d_2^d $ & 0.0396(311) & 0.0313(193) & 0.0230(104) & 0.0193(65) & \\
\end{tabular}

\caption{Moments of quark distributions in quenched QCD calculated at three values of
$\kappa$ and extrapolated linearly  to $\kappa_c$.}
\label{quenched_data}
\end{table}

\begin{table}[!ht]
\centering
\begin{tabular}{|l|c|ccccc|}
  & $ \kappa_c$=0.1585  & (0.1575) & 0.1570 & 0.1565 & 0.1560 & \\
\hline
\hline
$\langle x \rangle _u^{(b)}$  & 0.459(29) & 0.503(14) & 0.470(13) & 0.449(9) & 0.46
1(5) & \\
$\langle x \rangle _d^{(b)}$  & 0.190(17) & 0.222(9) & 0.207(8) & 0.207(4) & 0.214(4
) & \\
\hline
$\langle x \rangle _u^{(a)}$  & 0.462(174) & 0.412(93) & 0.411(69) & 0.439(70) & 0.
394(45) & \\
$\langle x \rangle _d^{(a)}$ & 0.178(83) & 0.168(37) & 0.161(33) & 0.216(37) & 0.17
0(20) & \\
\hline
$\langle x ^2 \rangle _u$  & 0.176(63) & 0.131(34) & 0.134(24) & 0.131(24) & 0.110(18
) & \\
$\langle x ^2 \rangle _d$& 0.0314(303) & 0.0328(151) & 0.0414(109) & 0.0503(102) 
& 0.0496(101) & \\
\hline
$\langle x^3 \rangle _u$  & 0.0685(392) & 0.0443(168) & 0.0521(152) & 0.0594(121) 
& 0.0466(113) & \\
$\langle x^3 \rangle _d$   & -0.00989(1529) & 0.0232(98) & 0.00789(549) & 0.0259(
63) & 0.0225(50) & \\
\hline
$\langle 1 \rangle _{\Delta u}$  & 0.860(69) & 0.741(37) & 0.880(28) & 0.975(24) &
0.936 (16) & \\
$\langle 1 \rangle _{\Delta d}$    & -0.171(43) & -0.214(25) & -0.214(18) & -0.248(15)
&  -0.254(9) & \\
\hline
$\langle x \rangle _{\Delta u}^{(b)}$ & 0.242(22) & 0.241(15) & 0.237(9) & 0.237(7) & 
0.235(6) & \\
$\langle x \rangle _{\Delta d}^{(b)}$ & -0.0290(129) & -0.0484(63) & -0.0460(52) & -
0.0605(38) & -0.0621(34) & \\
\hline
$\langle x \rangle _{\Delta u}^{(a)}$  & 0.254(111)& 0.205(83) & 0.196(43) & 0.217(41) 
& 0.171(31) & \\
$\langle x \rangle _{\Delta d}^{(a)}$  & -0.0546(863) & -0.0611(418) &
-0.0849(377)  & -0.0473(246) & -0.0745(194) & \\
\hline
$\langle x^2 \rangle _{\Delta u}$  & 0.116(42) & 0.0859(330) & 0.0673(161) &
0.0920(166 ) & 0.0483(114) & \\
$\langle x^2 \rangle _{\Delta d}$ & 0.00142(2515) & -0.0179(169) &
-0.0149(101) &
 -0.0157(95) & -0.0239(61) & \\
\hline
$\langle 1 \rangle _{\delta u}$  & 0.963(59) & 0.919(40) & 1.023(26) & 1.062(18) &
1.075 (13) & \\
$\langle 1 \rangle _{\delta d}$ & -0.202(36) & -0.239(27) & -0.238(15) &
-0.225(14) &
 -0.250(7) & \\
\hline
$\langle x \rangle _{\delta u}$   & 0.477(196)& 0.424(109) & 0.418(79) & 0.465(75) & 0
.405(48) & \\
$\langle x \rangle _{\delta d}$ & -0.144(68) & -0.0828(385) & -0.115(28) & -0.0565(24
8) & -0.0739(167) & \\
\hline
$ d_1^u $  & -1.318(55) & -1.036(44) & -1.032(26) & -0.957(15) & -0.
854(11) & \\
$ d_1^d $  & 0.278(35) & 0.269(30) & 0.239(15) & 0.200(13) & 0.197(6)
 & \\
\hline
$ d_2^u $  & -0.228(81) & -0.191(50) & -0.179(34) & -0.164(29) & -0.
147(17) & \\
$ d_2^d $  & 0.0765(310) & 0.0392(181) & 0.0462(134) & 0.0172(112) 
& 0.0154(59) & \\
\end{tabular}

\caption{Moments of quark distributions in full QCD at $\beta = 5.6$ using SESAM
configurations. Lattice measurements are shown  at four values of
$\kappa$, but to avoid finite volume effects at the lightest quark mass $(\kappa =
0.1575)$,  only values at  the three lowest  $\kappa$'s are extrapolated linearly  to
$\kappa_c$ .}
\label{SESAM_data}
\end{table}

\begin{table}[!ht]
\centering
\begin{tabular}{|l|c|cccc|}
 \hline 
 & $ \kappa_c$=0.16116 & 0.1604 & 0.1600 & 0.1596 & \\
\hline
\hline
$\langle x \rangle _u^{(b)}$ & 0.416(41) & 0.459(25) & 0.451(16) & 0.480(9) &
\\
$\langle x \rangle _d^{(b)}$ & 0.181(20) & 0.197(10) & 0.205(11) & 0.213(7) &
\\
\hline
$\langle x \rangle _u^{(a)}$ & 0.289(132) & 0.372(68) & 0.420(64) & 0.462(47)
& \\
$\langle x \rangle _d^{(a)}$ & 0.0430(523) & 0.119(26) & 0.182(29) &
0.209(20) & \\
\hline
$\langle x ^2 \rangle _u$ & 0.0191(429) & 0.0753(210) & 0.150(30) &
0.146(17) & \\
$\langle x ^2 \rangle _d$ & 0.0290(311) & 0.0435(171) & 0.0513(134) &
0.0590(94) & \\
\hline
$\langle x^3 \rangle _u$ & 0.0341(365) & 0.0401(201) & 0.0500(152) &
0.0509(112) & \\
$\langle x^3 \rangle _d$ & 0.0232(228) & 0.0229(138) & 0.00746(827) &
0.00975(583) & \\
\hline
$\langle 1 \rangle _{\Delta u}$ & 0.635(94) & 0.713(52) & 0.838(40) &
0.852(27) & \\
$\langle 1 \rangle _{\Delta d}$ & -0.279(48) & -0.263(25) & -0.280(22) &
-0.259(16) & \\
\hline
$\langle x \rangle _{\Delta u}^{(b)}$ & 0.180(34) & 0.204(20) & 0.209(13) &
0.223(9) & \\
$\langle x \rangle _{\Delta d}^{(b)}$ & -0.0597(168) & -0.0597(94) &
-0.0597(70) & -0.0597(48) & \\
\hline
$\langle x \rangle _{\Delta u}^{(a)}$ & 0.278(123) & 0.217(66) & 0.255(57) &
0.192(39) & \\
$\langle x \rangle _{\Delta d}^{(a)}$ & 0.119(61) & 0.0419(316) & -0.0147(365)
& -0.0459(201) & \\
\hline
$\langle x^2 \rangle _{\Delta u}$ & 0.156(50) & 0.0987(265) & 0.117(28) &
0.0598(153) & \\
$\langle x^2 \rangle _{\Delta d}$ & 0.00611(3014) & -0.0214(159) &
0.0230(162) & -0.0223(93) & \\
\hline
$\langle 1 \rangle _{\delta u}$ & 0.675(75) & 0.868(42) & 0.980(31) &
1.080(22) & \\
$\langle 1 \rangle _{\delta d}$  & -0.214(55) & -0.225(30) & -0.277(25) &
-0.265(14) & \\
\hline
$\langle x \rangle _{\delta u}$ & 0.251(138) & 0.376(71) & 0.421(76) &
0.501(46) & \\
$\langle x \rangle _{\delta d}$ & 0.00912(7595) & 0.0204(412) & -0.106(31) &
-0.0527(253) & \\
\hline
$ d_1^u $ & -1.261(89) & -1.197(56) & -1.068(32) & -1.046(21) & \\
$ d_1^d $ & 0.374(67) & 0.302(41) & 0.301(27) & 0.256(13) & \\
\hline
$ d_2^u $ & -0.264(77) & -0.236(42) & -0.203(40) & -0.196(21) & \\
$ d_2^d $ & 0.0191(402) & 0.00394(2320) & 0.0501(172) & 0.0266(100) & \\
\hline
\end{tabular}
\caption{Moments of quark distributions in full QCD at $\beta = 5.5$ using SCRI
configurations extrapolated linearly  to
$\kappa_c$. }
\label{SCRI_data}

\end{table}

\begin{table}[!ht]
\centering
\begin{tabular}{|l|c|ccc|}
 \hline 
 & $ \kappa_c$=0.1266 & 0.1235 & 0.1246 & \\
\hline
\hline
$\langle x \rangle _u^{(b)}$ & 0.565(61) & 0.581(8) & 0.575(21) & \\
$\langle x \rangle _d^{(b)}$ & 0.238(29) & 0.265(5) & 0.255(10) & \\
\hline
$\langle x \rangle _u^{(a)}$ & 0.314(189) & 0.555(74) & 0.468(49) & \\
$\langle x \rangle _d^{(a)}$ & 0.119(89) & 0.247(33) & 0.201(24) & \\
\hline
$\langle x ^2 \rangle _u$ & 0.146(78) & 0.169(29) & 0.161(21) & \\
$\langle x ^2 \rangle _d$ & 0.0316(334) & 0.0716(123) & 0.0572(91) & \\
\hline
$\langle x^3 \rangle _u$ & 0.0517(337) & 0.0761(130) & 0.0673(88) & \\
$\langle x^3 \rangle _d$ & 0.00911(1390) & 0.0309(55) & 0.0231(35) & \\
\hline
$\langle 1 \rangle _{\Delta u}$ & 0.585(82) & 0.818(17) & 0.734(28) & \\
$\langle 1 \rangle _{\Delta d}$ & -0.298(46) & -0.210(8) & -0.242(16) & \\
\hline
$\langle x \rangle _{\Delta u}^{(b)}$ & 0.118(46) & 0.114(17) & 0.115(13) & \\
$\langle x \rangle _{\Delta d}^{(b)}$ & -0.0120(242) & -0.0230(57) &
-0.0191(79) & \\
\hline
$\langle x \rangle _{\Delta u}^{(a)}$ & 0.279(30) & 0.316(8) & 0.303(9) & \\
$\langle x \rangle _{\Delta d}^{(a)}$ & -0.0763(188) & -0.0650(46) &
-0.0691(61) & \\
\hline
$\langle x^2 \rangle _{\Delta u}$ & 0.236(117) & 0.274(40) & 0.261(34) & \\
$\langle x^2 \rangle _{\Delta d}$ & -0.0650(782) & -0.0534(200) &
-0.0576(251) & \\
\hline
$\langle 1 \rangle _{\delta u}$ & 0.768(65) & 0.951(13) & 0.885(22) & \\
$\langle 1 \rangle _{\delta d}$ & -0.289(49) & -0.234(7) & -0.254(17) & \\
\hline
$\langle x \rangle _{\delta u}$ & 0.451(265) & 0.702(96) & 0.611(73) & \\
$\langle x \rangle _{\delta d}$ & 0.0153(936) & -0.137(27) & -0.0820(288) &
\\
\hline
$ d_1^u $ & -0.101(7) & 0.155(2) & 0.0628(22) & \\
$ d_1^d $ & 0.00915(595) & -0.0398(12) & -0.0221(20) & \\
\hline
$ d_2^u $ & -0.0462(220) & 0.0619(95) & 0.0229(51) & \\
$ d_2^d $ & 0.00673(1288) & -0.0164(34) & -0.00808(409) & \\
\hline
\end{tabular}
\caption{Moments of quark distributions calculated with cooled configurations, as
described in the text, to eliminate most of the gluon degrees of freedom except
instantons. Lattice measurements  at  $\kappa = 0.1235 $ and $ 0.1246 $ are
extrapolated linearly  to
$\kappa_c$.}
\label{cooled _data}
\end{table}

\begin{table}[hbt]

\begin{center}
\begin{footnotesize}
\begin{tabular}{lrrrrrr}
\hline
Connected & QCDSF & QCDSF & Wuppertal &Quenched &Full QCD 
& \multicolumn{1}{c}{Phenomenology} \\
 M. E.&  & ($a=0$) &  &  & (3 pts) & \multicolumn{1}{c}{($q\pm \bar q$)} \\  
\hline
$ \langle x \rangle_u $      & $0.452(26)$ &   &   & $0.454(29)$ & $0.459(29)$ &  \\
$\langle x \rangle_d $      & $0.189(12)$ &   &   & $0.203(14)$ & $0.190(17)$ &  \\
$\langle x \rangle_{u-d}$    & $0.263(17)$ &   &   & $0.251(18)$ & $0.269(23)$ &
$0.154 (3)$ \\ 
$\langle x^2 \rangle_u $    & $0.104(20)$ &   &   & $0.119(61)$ & $0.176(63)$ &  \\
$ \langle x^2 \rangle_d $    & $0.037(10)$ &   &   & $0.029(32)$ & $0.031(30)$ & 
\\
$\langle x^2 \rangle_{u-d} $ & $0.067(22)$& &  & $0.090(68)$ &  $0.145(69) $
& $0.055 (1) $\\
$ \langle x^3 \rangle_u $    & $0.022(11)$ &   &   & $0.037(36)$ & $0.069(39)$ &
 \\
$ \langle x^3 \rangle_d $    & $-0.001(7)$ &   &   & $0.009(18)$ & $-0.010(15)$
 \\
$\langle x^3 \rangle_{u-d} $ & $0.023(13)$& &  & $0.028(49)$ &  $0.078(41) $
& $0.023 (1) $\\
$ \langle 1\rangle_{\Delta u}$ & $0.830(70)$ & $0.889(29)$ & $0.816(20)$
& $0.888(80)$ & $0.860(69)$ &  \\
$\langle 1\rangle_{\Delta d}$ & $-0.244(22)$ & $-0.236(27)$ & $-0.237(9)$
& $-0.241(58)$ & $-0.171(43)$ &  \\
$ \langle 1\rangle_{\Delta u -\Delta d} $ & $1.074(90)$ & $1.14(3)$ & $1.053(27)$  
& $1.129(98)$ & $1.031(81)$ & $1.248 (2)$ \\
$\langle x \rangle_{\Delta u} $ & $0.198(8)$ &   &  & $0.215(25)$ & $0.242(22)$ & \\
$\langle x \rangle_{\Delta d}$ & $-0.048(3)$ &   &  & $-0.054(16)$ &
$-0.029(13)$ & \\
$\langle x \rangle_{\Delta u -\Delta d}$ & $0.246(9)$ &      &  & $0.269(29)$ &
$0.271(25)$ & $0.196 (9)$ \\
$ \langle x ^2\rangle_{\Delta u} $ & $0.04(2)$ &      &  & $0.027(60)$ &
$0.116(42)$ & \\
$ \langle x ^2\rangle_{\Delta d} $ & $-0.012(6)$ &      &  & $-0.003(25)$ &
$0.001(25)$ &\\
$ \langle x ^2\rangle_{\Delta u - \Delta d} $ & $0.05(2)$ &      &  & $0.030(65)$
&
$0.115(49)$ & $0.061 (6)$ \\
$ \delta u_c $ & $0.93(3)$ & $0.980(30)$ &   & $1.01(8)$ & $0.963(59)$ &  \\
$ \delta d_c $ & $-0.20(2)$ & $-0.234(17)$ &   & $-0.20(5)$ & $-0.202(36)$ &  \\
$ d_2^u $ & $-0.206(18)$ &     &   & $-0.233(86)$ & $-0.228(81)$ &  \\
$ d_2^d $ & $-0.035(6)$ &     &   & $0.040(31)$ & $0.077(31)$ &  \\
\hline
\end{tabular}
\end{footnotesize}
\end{center}
\caption{Comparison of linear extrapolations of  full QCD and quenched
results with other lattice calculations and phenomenology at  $4$ GeV$^2$ in
the
$\overline{MS}$ scheme. The first column shows quenched results by the QCDSF
collaboration at $\beta=6.0$
\protect\cite{bib-qcdsf-main,bib-qcdsf-a1,bib-qcdsf-spin,bib-qcdsf-d2,Gockeler:2000ja}
and the second column shows extrapolation of several moments to the
continuum limit\protect\cite{bib-qcdsf-tensor2}. The third column shows full
QCD results calculated using a different method with the same SESAM
configurations we have used\protect\cite{Gusken:1999xy}. The quenched and
full QCD results calculated in this work are shown in the fourth and fifth
columns. Flavor non-singlet moments
$\langle x^n\rangle$ of
$q (x) + (-1)^{n+1}\bar q(x)$, $\Delta q (x) + (-1)^{n}\Delta\bar q(x)$, and $\delta q
(x) + (-1)^{n+1}\delta\bar q(x)$ are tabulated in the final column. Phenomenological
unpolarized distributions are calculated from
refs.~\protect\cite{bib-cteq,bib-grv,bib-mrs} and  polarized distributions are
calculated from refs.~\protect\cite{bib-grsv,bib-gs} with error estimates as
described in the text. }
\label{tab-summary}
\end{table}

\newpage

\end{document}